\documentclass[10.5pt]{iopart}

\usepackage{iopams,amsfonts,amssymb,amsthm} 

\pdfoutput=1
\usepackage{etex}
\usepackage{cite}
\usepackage{geometry}                
\usepackage{graphicx}
\usepackage{rawfonts}
\usepackage{amssymb}
\usepackage{epstopdf}
\usepackage{fullpage}
\usepackage{color}
\usepackage[usenames,dvipsnames,svgnames]{xcolor}
\usepackage[colorlinks=true,citecolor=blue,linkcolor=red,urlcolor=blue]{hyperref}
\usepackage[font=footnotesize, labelfont=bf, margin=0.5cm]{caption}
\usepackage[labelformat=simple]{subcaption}
\usepackage{rawfonts}
\usepackage{etex}
\usepackage{tikz}
\usepackage{multirow}
\usepackage{rotating}
\usepackage{lscape}
\usepackage{bbm}
\usepackage{latexsym}
\usepackage[flushleft]{threeparttable}

\input prepictex
\input pictex
\input postpictex

\def\2;{\;\;}

\def\eps{\epsilon}

\def\IntZ{{\mathbb Z}}

\def\Ref#1{(\ref{#1})}

\def\x#1{{x_#1}}

\def\binom#1#2{{{#1}\choose{#2}}}
\def\Bi#1#2{{\binom{#1}{#2}}}

\def\Sfrac#1#2{\hbox{\large $\frac{#1}{#2}$}}

\def\LB{\left(}         \def\RB{\right)}

\def\lcl{\left\lceil}  \def\rcl{\right\rceil}
\def\lfl{{\!}\left\lfloor} \def\rfl{\right\rfloor{\!}}
\def\LC{\left\{}       \def\RC{\right\}}

\def\svv{{\;\hbox{$|$}\;}}

\def\tiny#1{\scalebox{0.75}{#1}}

\definecolor{blue}{rgb}{0,0.18,0.39}
\definecolor{RoyalBlue}{rgb}{0,0.2,0.7}

\definecolor{Maroon}{cmyk}{0,0.87,0.68,0.62}
\definecolor{Brown}{rgb}{0.7,0.3,0}
\definecolor{Navy}{rgb}{0.3,0.0,0.4}
\definecolor{Red}{cmyk}{0,1,1,0}
\definecolor{BrickRed}{cmyk}{0.16,0.89,0.61,0.02}
\definecolor{DarkRed}{cmyk}{0,1,1,0.5}
\definecolor{DarkBlue}{cmyk}{1,1,0,0.2}
\definecolor{DarkGreen}{cmyk}{1,0,1,0.4}
\definecolor{Green}{cmyk}{1,0,1,0}
\definecolor{DarkBrown}{cmyk}{0,0.81,1,0.6}
\definecolor{OrangeRed}{cmyk}{0,1,0.87,0}
\definecolor{RedOrange}{cmyk}{0,0.77,0.87,0}
\definecolor{Orange}{cmyk}{0,0.61,0.87,0}
\definecolor{Offwhite}{rgb}{.8,0.9,.8}
\definecolor{Offwhite2}{cmyk}{.04,.02,.01,0}
\definecolor{Tan}{rgb}{0.82,0.70,0.55}
\definecolor{Blue}{rgb}{0,0,1}
\definecolor{RoyalBlue}{rgb}{0.25,0.41,0.88}
\definecolor{Sepia}{rgb}{0.37,0.14,0.07}
\definecolor{myblue}{cmyk}{0.025,0.05,0,0}
\definecolor{Mahogany}{cmyk}{0.18,0.87,1,0.08}

\definecolor{green1}{cmyk}{0.25,0,0.76,0}
\definecolor{green2}{cmyk}{0.25,0,0.76,0.07}
\definecolor{green3}{cmyk}{0.25,0,0.76,0.20}
\definecolor{green4}{cmyk}{0.25,0,0.75,0.30}
\definecolor{green5}{cmyk}{0.25,0,0.75,0.40}
\definecolor{green6}{cmyk}{0.25,0,0.75,0.50}

\definecolor{B02}{cmyk}{0,0.14,0.22,0.12}
\definecolor{B03}{cmyk}{0,0.16,0.26,0.16}
\definecolor{B04}{cmyk}{0,0.19,0.28,0.19}
\definecolor{B05}{cmyk}{0,0.25,0.32,0.25}
\definecolor{B06}{cmyk}{0,0.31,0.36,0.31}
\definecolor{B07}{cmyk}{0,0.37,0.40,0.37}
\definecolor{B08}{cmyk}{0,0.46,0.46,0.46}
\definecolor{B09}{cmyk}{0,0.55,0.52,0.54}
\definecolor{B10}{cmyk}{0,0.69,0.61,0.62}
\definecolor{B11}{cmyk}{0,0.78,0.70,0.68}
\definecolor{B12}{cmyk}{0,0.93,0.85,0.60}
\definecolor{B13}{cmyk}{0.25,1,0.6,0.50}
\definecolor{B14}{cmyk}{0.5,1,0.30,0.40}
\definecolor{B15}{cmyk}{0.75,1,0,0.30}

\definecolor{C02}{cmyk}{0,0.22,0.14,0.12}
\definecolor{C03}{cmyk}{0,0.26,0.16,0.16}
\definecolor{C04}{cmyk}{0,0.28,0.19,0.19}
\definecolor{C05}{cmyk}{0,0.32,0.25,0.25}
\definecolor{C06}{cmyk}{0,0.36,0.31,0.31}
\definecolor{C07}{cmyk}{0,0.40,0.37,0.37}
\definecolor{C08}{cmyk}{0,0.46,0.46,0.46}
\definecolor{C09}{cmyk}{0,0.52,0.55,0.54}
\definecolor{C10}{cmyk}{0,0.61,0.69,0.62}
\definecolor{C11}{cmyk}{0,0.70,0.78,0.68}
\definecolor{C12}{cmyk}{0,0.85,0.93,0.60}
\definecolor{C13}{cmyk}{0.25,0.60,1,0.50}
\definecolor{C14}{cmyk}{0.5,0.30,1,0.40}
\definecolor{C15}{cmyk}{0.75,0,1,0.30}

\usepackage[font=footnotesize,labelfont=sf]{caption}

\newtheorem{theorem}{Theorem}
\newtheorem{lemma}{Lemma}
\newtheorem{corollary}{Corollary}

\newtheorem{conjecture}{Conjecture}

\def\ra{\hbox{\beginpicture \plot 1 5 1 1 5 1 / \endpicture}}
\def\dps{\displaystyle}

\begin{document}

\title[Walk under a piston]{\textsf{The escape transition in a self-avoiding walk model of linear polymers}}

\author{EJ Janse van Rensburg$^1$}

\address{$^1$Department of Mathematics and Statistics, 
York University, Toronto, Ontario M3J~1P3, Canada}
\ead{$\ddagger$\href{mailto:rensburg@yorku.ca}{rensburg@yorku.ca}}
\vspace{10pt}
\begin{indented}
\item[]\today
\end{indented}

\begin{abstract}
A linear polymer grafted to a hard wall and underneath an AFM tip can be modelled
in a lattice as a grafted lattice polymer (or self-avoiding walk) compressed 
underneath a piston approaching the wall.  As the piston approaches the wall the 
increasingly confined polymer escapes from the confined region to explore 
conformations beside the piston.  This conformational change is believed to be
a phase transition in the thermodynamic limit, and has been argued to be first 
order, based on numerical results in reference \cite{HBKS07}.  In this paper
a lattice self-avoiding walk model of the escape transition is constructed.
It is proven that this model has a critical point in the thermodynamic limit
corresponding to the escape transition of compressed grafted linear polymers.  
This result relies on the analysis of ballistic self-avoiding walks 
in slits and slabs in the square and cubic lattices.  Additionally, numerical estimates
of the location of the escape transition critical point is reported based on 
Monte Carlo simulations of self-avoiding walks in slits and in slabs. 
\end{abstract}

%
\vspace{0pc}


\section{Introduction}

The properties of polymers grafted to hard walls or interfaces, or in confined geometries, 
are of significant interest in polymer physics \cite{deG79}.  These properties underlie 
important applications of polymers, including the stabilization of colloids 
\cite{LTR78,F85,WP86,NDA97}, the in vivo adsorption and delivery of drugs using polymer 
coatings on medical devices, such as stents \cite{CVZ10,JM12,SYS15}, the behaviour 
of biopolymers at cell membranes \cite{H18}, or the interaction of grafted polymers 
and small particles \cite{SWP96}, amongst many other examples.  

Confinement and manipulation of single polymer molecules have become possible
using atomic force microscopy (AFM).   Confining a polymer which is grafted to 
hard wall by an approaching tip of the atomic force microscope reduces the 
conformational degrees of freedom of the polymer, and it may undergo an 
``escape'' transition where part of it escapes from underneath the tip to 
explore conformations in the region beside the tip \cite{GWS97}.  

A lattice model of a grafted linear polymer being compressed by the AFM tip is shown 
in figure \ref{f1}.  A lattice self-avoiding walk is grafted at the origin in the hard 
wall (or ``anvil''), and explores its conformations in a space $\Sigma_w$ above 
the anvil and below or outside the AFM tip (the ``piston'').  If the piston is far 
above the anvil, then $\Sigma_w$ is large, and the walk explores it conformations 
primarily below the piston.  As the piston approaches the anvil, the conformational 
degrees of freedom of the walk is reduced, and the walk eventually escapes 
from the space below the piston.  In this case the walk is stretch to the boundary 
of the piston, and its remaining part (or ``tail'') explores conformations primarily 
outside the piston (rather than below it).

The model in figure \ref{f1} is a lattice version of models examined in a series of
excellent papers \cite{GWS97,MYB99, HBKS07,PMEB13} exploring the scaling and 
transition in linear and star polymers compressed by a piston.  Additional numerical
results on star polymers can be found in references \cite{S00,RC15}. These studies
are of bead-spring models \cite{MYB99}, lattice models \cite{HBKS07}, and 
molecular dynamics simulations \cite{PMEB13}.  While an escape transition is
not established rigorously in these models, there are  ample numerical 
evidence of such a transition.  In references \cite{GWS97,MYB99} a theoretical 
approach using phenomenological arguments based on a ``blob'' analysis 
(see, for example, \cite{deG79}) of the confined polymer is pursued. 
The analysis in reference \cite{GWS97} proceeds by considering the free energy as 
a function of the separation between the anvil and the piston, while reference
\cite{MYB99} proceeds by considering the escape transition as a function
of the force $f$ exerted on the piston by the polymer (this force is conjugate to the
separation between the anvil and the piston).  In these references the phenomenological 
blob analysis and numerical simulations using a bead-spring and other models 
show convincing evidence of an escape transition in two dimensions. However,
the order of the transition is still unresolved \cite{HBKS07}.

\subsection{Lattice models and main results}
\label{s2}

\begin{figure}[h]
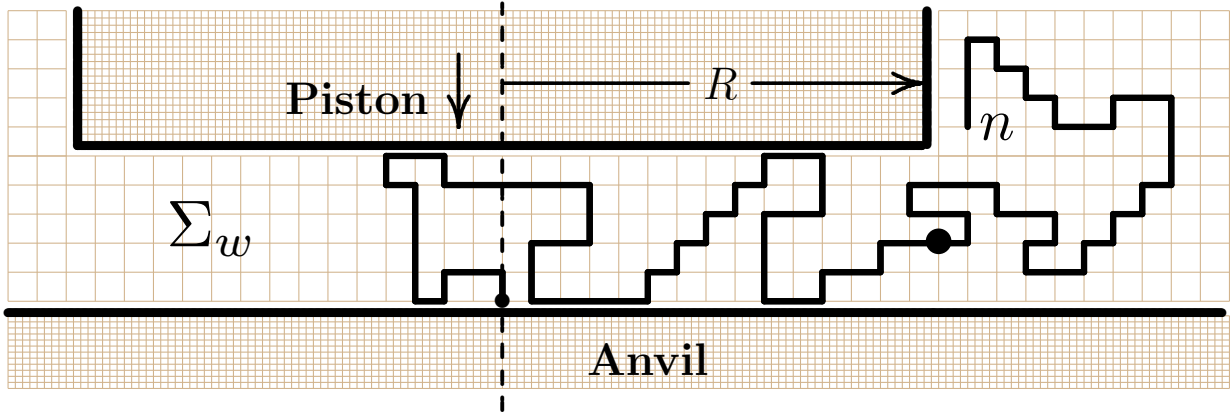


\beginpicture
\setcoordinatesystem units <1.1pt,1.1pt>
\setplotarea x from 0 to 300, y from -30 to 100
\setplotsymbol (.) \color{Tan}
\setplotarea x from -120 to 300, y from  0 to 50
\grid 42 5 
\setplotarea x from -120 to -100, y from 50 to 100
\grid 2 5 
\setplotarea x from 200 to 300, y from 50 to 100
\grid 10 5
\setplotarea x from -120 to 300, y from -30 to -5
\grid 168 12
\setplotarea x from -95 to 195, y from 55 to 100
\grid 119 18

\setplotarea x from -100 to 300, y from -35 to 105

\setplotsymbol ({\footnotesize$\bullet$}) \color{Black}
\multiput {\Large$\bullet$} at 50 0  /
\plot -120 -4 200 -4 300 -4 /
\plot -96 100 -96 53.5 196 53.5 196 100 /

\setplotsymbol ({$\scalebox{2.25}{.}$})

\plot 50 0 50 10 30 10 30 0 20 0 20 40 10 40 10 50 30 50 30 40
80 40 80 20 60 20 60 0 100 0 100 10 110 10 110 20 120 20 120 30
130 30 130 40 140 40 140 50 160 50 160 30 140 30 140 0 160 0
160 10 180 10 180 20 200 20 /
\put {$\scalebox{2.5}{\hbox{$\bullet$}}$} at 200 20 
\plot 200 20 210 20 210 30 190 30 190 40 220 40 220 30 240 30 
240 20 230 20 230 10 250 10 250 20 260 20 260 30 270 30 270 40
280 40 280 70 260 70 260 60 240 60 240 70 230 70 230 80 220 80
220 90 210 90 210 60 /

\put {\LARGE \bf Piston} at 0 70
\put {\LARGE \bf Anvil} at 100 -20
\multiput {\scalebox{2.5}{$\Sigma_w$}} at -50 25  /

\setplotsymbol ({$\scalebox{1.25}{.}$})
\arrow <12pt> [.2,.67] from 35 85 to 35 60

\plot 50 75 114 75 / 
\arrow <12pt> [.2,.67] from 136 75 to 195 75 
\put {\LARGE $R$} at 125 75
\put {\hbox{\scalebox{2.25}{$n$}}} at 220 60

\setdashes <5pt> \plot 50 -37.5 50 102.5 /

\color{black} \normalcolor
\endpicture
\caption{A lattice model of a linear polymer grafted to a surface (the ``anvil'') and
being squeezed by a piston approaching the anvil.  If the polymer is long, then part
of it may escape from the confining space underneath the piston into the bulk
region beside the piston.}
\label{f1}
\end{figure}

\begin{figure}[h]
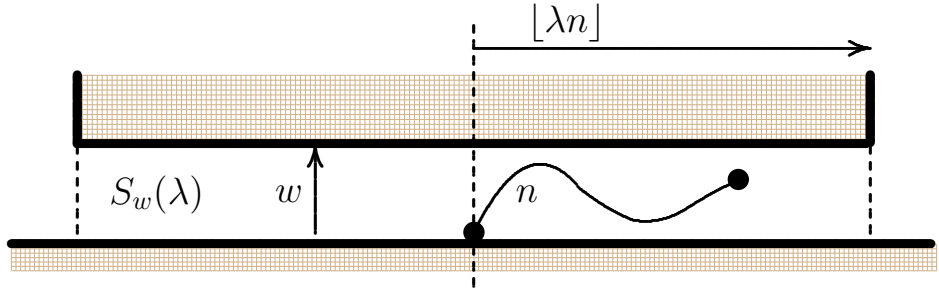


\beginpicture
\setcoordinatesystem units <1pt,1pt>
\setplotarea x from -250 to 150, y from -20 to 75
\color{Tan}
\setplotarea x from -175 to 175, y from -14 to -4
\grid 200 6
\setplotarea x from -150 to 150, y from 34 to 60
\grid 160 14

\color{black}
\setplotarea x from -150 to 150, y from -20 to 75
\setplotsymbol ({\footnotesize$\bullet$})
\plot -175 -4 175 -4 /
\plot -150 60 -150 34 150 34 150 60 /
\setplotsymbol ({$\cdot$})

\setquadratic 
\plot 0 0  20 25  40 17  50 10  60 5  70 5  80 9  90 15 100 20 /
\setlinear
\multiput {\huge$\bullet$} at 0 0 100 20 /

\arrow <8pt> [.2,.67] from -60 0 to -60 33 \put {\Large$w$} at -70 15
\arrow <8pt> [.2,.67] from 0 70 to 150 70   \put {\Large$\lfl\lambda n\rfl$} at 35 80
\put {\Large$n$} at 20 15
\put {\Large$S_w(\lambda)$} at -120 15

\setdashes <3pt> \plot 0 -20 0 80 /
\plot -150 0 -150 33 /  \plot 150 0 150 33 / 

\color{black} \normalcolor
\endpicture

\caption{Schematic of a walk of length $n$ being squeezed by a piston of
radius $\lfl \lambda n\rfl$ with $\lambda>1$. In this case the walk is confined to the
slit or slab $S_w(\lambda)$ underneath the piston and above the anvil.
Notice that $S_w(\lambda)$ does not extend beyond the edge of the piston.}
\label{f2}
\end{figure}

Let $c_n$ be the number of self-avoiding walks from the origin in the
$d$-dimensional hypercubic lattice. Then the \textit{growth constant} $\mu_d$ of
the self-avoiding walk is defined by
\begin{equation}
\log \mu_d = \lim_{n\to\infty} \Sfrac{1}{n} \log c_n .
\label{1}
\end{equation}
The growth constant has been estimated to high accurancy in the square and
cubic lattices, namely
\begin{equation}
\mu_d =
\cases{
2.63815853032790(3) , & \hbox{if $d=2$ (square lattice) \cite{JSG16}}; \cr
4.684039931(27), & \hbox{if $d=3$ (cubic lattice) \cite{C13}} .
}
\label{2}
\end{equation}
If the walk is confined by boundaries in the lattice, then the value of the
growth constant may change, and this is in particular the case if the walk
is confined by a piston when it is grafted to a hard wall (anvil).  In figure \ref{f1} a square 
lattice model of a piston of radius $R$ compressing a self-avoiding walk against 
an anvil is shown.  As the piston approaches the anvil, the polymer is confined 
to a region $\Sigma_w$, which consists of the space underneath and beside 
the piston and above the anvil.  The thermodynamic limit in this model is taken 
by fixing the ratio of the piston radius $R$ to the length of the walk $n$, and 
then taking $n\to\infty$ with $R/n$ fixed.  In the lattice geometry this is 
achieved by putting $R=\lfl\lambda n \rfl$, as illustrated in figure \ref{f2}.  
In this paper it is shown that, in the thermodynamic limit, there 
exists a phase transition in this model, in both the square and cubic lattices.

In the lattice model in figure \ref{f1} the origin is located on the anvil 
centered underneath the piston.  In the square lattice the piston is a rectangle
with vertical bisector running through the origin.  In the cubic lattice, the piston
may be assumed to have a square or circular horisontal projection onto the
anvil, and its vertical symmetry axis runs through the origin.  The linear polymer 
is a self-avoiding walk of length $n$ from the origin (or grafted at the origin), and 
confined to explore conformations in $\Sigma_w$. 

The vertical distance between the anvil and the piston is $w$, and the 
\textit{radius} of the piston is the length of the shortest self-avoiding walk
from the origin to a vertex underneath the edge of the piston.  If a piston
has radius $\lfl \lambda n \rfl$, then the number of self-avoiding 
walks of length $n$ from the origin in $\Sigma_w$ is denoted by $w_n(\lambda)$.  
The free energy of this model, per unit length, is given by
\begin{equation}
\rho_n(\lambda) = \Sfrac{1}{n} \log w_n(\lambda) .
\label{e1}
\end{equation}
The limit of $\rho_n(\lambda)$ as $n\to\infty$ is the \textit{limiting free energy} 
of the model, and should be compared to $\log \mu_d$ in equation \Ref{1}.
In this paper we show that for a range of values of $\lambda$ the limit as $n\to\infty$
of the function in equation \Ref{e1} exists.  The main result is theorem \ref{T1}.

\begin{theorem}
Define $\lambda_0=1/(w{+}1)$ in the square lattice, and $\lambda_0=0$ in the
cubic lattice. Then there exists a $\lambda_1 \in [\lambda_0,1)$ such that the
limiting free energy of a walk from the origin in $\Sigma_w$ is given by
$$ \rho^{(w)} (\lambda) = \lim_{n\to\infty} \Sfrac{1}{n} \log w_n(\lambda) $$
for every $\lambda \in [\lambda_1,1]$ in the square lattice, or in the cubic lattice. \qed
\label{T1}
\end{theorem}

In figure \ref{f2} the region underneath the piston and above the anvil is 
denoted by $S_w(\lambda)$.   Notice that $S_w(\lambda)$ does not extend 
beyond the edge of the piston, and so is a finite part of the square lattice 
(that is, $S_w(\lambda) \subset \Sigma_w$).  $S_w(\lambda)$ is similarly 
defined in the cubic lattice, underneath the piston, and above the anvil, and 
it also does not extend beyond the boundary of the piston in any direction.  
In the square lattice $S_w(\lambda)$ is a \textit{slit} of length 
$2\lfl \lambda n \rfl$, and in the cubic lattice is a \textit{slab} of radius 
$\lfl \lambda n \rfl$.

The radius of the piston is $\lfl \lambda n \rfl$, so that if $\lambda > 1$, then 
a walk of length $n$ is entirely confined to $S_w(\lambda)$.  This is 
shown schematically in figure \ref{f2}.   In the limit as $n\to\infty$, $S_w(\lambda)$
becomes a slit or a slab of infinite extent denoted by $S_w(\infty) \equiv S_w$, 
and the walk is confined to it, even as the limit $n\to\infty$ is taken.  

Let $c_n^{(w)}$ be the number of self-avoiding walks from the origin in 
$S_w(\infty)$ of height $w\geq 0$.  It is known that the limit
\begin{equation}
\lim_{n\to\infty} \Sfrac{1}{n} \log c_n^{(w)} = \log \mu_w^{(d)}
\label{e2}
\end{equation}
exists in the square and cubic lattices \cite{W83}.  In the case that 
$\lambda \geq 1$, $w_n(\lambda) = c_n^{(w)}$ for all $n\geq 0$,
since the piston is wide enough to confine all the conformations of a walk of length 
$n$ to $S_w(\lambda)$.  This shows that
\begin{equation}
\rho^{(w)} (\lambda) = \log \mu_w^{(d)},\qquad\hbox{if $\lambda \geq 1$} .
\label{e3}
\end{equation}
Moreover, $\rho^{(w)}(\lambda) \to \log \mu_d$ as $w\to\infty$ and $\lambda \geq 1$
\cite{WS91}.

\begin{figure}[h]
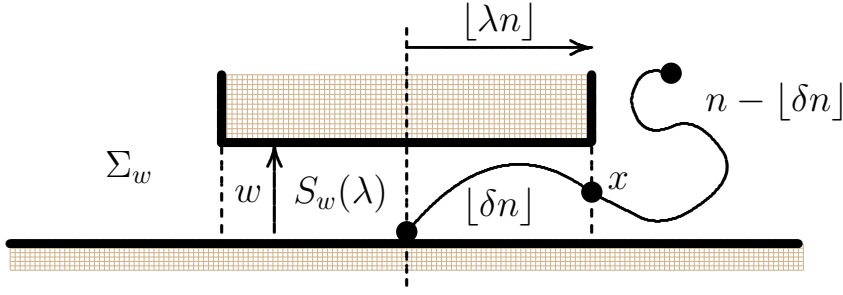


\beginpicture
\setcoordinatesystem units <1pt,1pt>
\setplotarea x from -200 to 150, y from -20 to 75
\color{Tan}
\setplotarea x from -150 to 150, y from -14 to -4
\grid 160 6
\setplotarea x from -70 to 70, y from 34 to 60
\grid 72 14

\color{black}
\setplotarea x from -150 to 150, y from -20 to 75
\setplotsymbol ({\footnotesize$\bullet$})
\plot -150 -4 150 -4 /
\plot -70 60 -70 34 70 34 70 60 /
\setplotsymbol ({$\cdot$})

\setquadratic 
\plot 0 0 35 25 70 15 /
\plot 70 15  80 10 90 5 100 5 110 10 120 20 120 30 110 40 100 40 
90 40 85 50 90 60 100 60 /
\setlinear
\multiput {\huge$\bullet$} at 0 0 100 60 70 15 /

\arrow <8pt> [.2,.67] from -50 0 to -50 33 \put {\Large$w$} at -60 15
\arrow <8pt> [.2,.67] from 0 70 to 70 70   \put {\Large$\lfl\lambda n\rfl$} at 35 80
\put {\Large$\lfl \delta n\rfl$} at 35 10
\put {\Large$n-\lfl\delta n\rfl$} at 140 50
\put {\Large$S_w(\lambda)$} at -25 15
\put {\Large$\Sigma_w$} at -105 25
\put {\Large$x$} at 80 20
\setdashes <2.5pt> \plot 0 -20 0 80 /

\setdashes <3pt>
\plot -70 0 -70 33 /  \plot 70 0 70 33 / 

\color{black} \normalcolor
\endpicture

\caption{Schematic of a walk of length $n$ escaping from $S_w(\lambda)$ underneath
the piston.  In this case the piston has radius $\lfl \lambda n \rfl$, and $\lambda<1$. 
The walk exits the slit or slab $S_w(\lambda)$ for the first time at $x$ and its 
part from the origin to $x$ has length $\lfl \delta n\rfl$ and is confined 
to $S_w(\lambda)$.  The remaining part of length $n-\lfl \delta n \rfl$ starts
at $x$ and may reenter and reexist $S_w(\lambda)$. Clearly,
 $\lambda \leq \delta \leq 1$.}
\label{f3}
\end{figure}

On the other hand, if $0\leq \lambda < 1$, then the walk may be partially inside
$S_w(\lambda)$ and then escape into the bulk regime outside $S_w(\lambda)$, as 
illustrated schematically in figure \ref{f3}.  If the walk escapes into the bulk as 
shown,  then it has a first part of length $\lfl \delta n \rfl$ from the origin to its 
first vertex underneath the edge of the piston before it steps outside $S_w(\lambda)$.
The remaining part of the walk has length $n{-}\lfl \delta n \rfl$ and explores
its conformations in $\Sigma_w$ (that is, it may reenter $S_w(\lambda)$).
Notice that $\lambda \leq \delta < 1$ in figure \ref{f3}.   In section 3 we show
there exists a $\lambda_c$ such that $\rho^{(w)}(\lambda) = \log \mu_w^{(d)}$ 
if $\lambda > \lambda_c$, and $\rho^{(w)}(\lambda) > \log \mu_w^{(d)}$ 
if $\lambda < \lambda_c$.  That is, $\rho^{(w)}(\lambda)$ has a non-analytic 
point at $\lambda_c \in [0,1)$.  This critical point corresponds to the escape transition 
of the walk.

This paper is organised as follows.  In section \ref{S2} models of ballistic walks in slits
and slabs in the square and cubic lattice are examined.   Existance of a thermodynamic
limit is proven in these cases using unfolded loops and walks in a slit or
in a slab.  These results are then used in section \ref{S3} to examine the full
model of walks underneath the piston and the escape transition.  Existence
of a critical point $\lambda_c$ is established, and a lower bound on it is
proven, namely
\begin{equation}
\lambda_c \geq \frac{\log (\mu_d/\mu_w^{(d)})}{\log \mu_d} .
\end{equation}
In addition to these results, numerical simulations of walks in a slit or 
slab using the PERM algorithm \cite{G97} in its flat histogram \cite{PK04} version, 
and with a parallel implementation \cite{CJvR20}, were done to determine the
free energy of ballistic walks in a slit or a slab.  Combining these 
results with the expressions for the free energy of walks underneath a piston
gives numerical approximations of $\lambda_c$, as shown in table \ref{tafel2}.

\section{Ballistic self-avoiding walks in slits and slabs}
\label{S2}

Denote the coordinates of vertices $v\in\IntZ^d$ in the hypercubic lattice
by $(\x1(v),\x2(v),\cdots,\x{d}(v))$ and recall that $S_w\equiv S_w(\infty)$ so that  
\begin{equation}
S_w = \{v\in\IntZ^d \svv 0 \leq \x{d}(v) \leq w \} .
\label{e4}
\end{equation}
The \textit{height} of $S_w$ is $w$.  As before, the number of self-avoiding
walks of length $n$ from the origin in $S_w$ is denoted by $c_n^{(w)}$ and the
growth constant $\mu_w^{(d)}$ of these walks is given in equation \Ref{e2}.

Generally 
$\log \mu_w^{(d)} < \log \mu_{w{+}1}^{(d)}$ \cite{SSW88A,SS98} and 
$\lim_{w\to\infty} \log \mu_w^{(d)} = \log \mu_d$ where $\mu_d$ is the
growth constant of the self-avoiding walk in $d$ dimensions (see
reference \cite{WS91} for more results and references 
\cite{CW87,WW87,WS91,WS92,JvROW06} for additional results and in 
particular lemma 8.18 and theorem 8.19 in reference \cite{JvR15}).  In the square
lattice $\log \mu_0^{(2)} = 0$ and $\log \mu_1^{(2)}>0$ while in the cubic 
lattice $\log \mu_0^{(3)} = \log \mu_2$.

If $w\to\infty$, then $\mu_w^{(d)} \to \mu_d$, the growth constant of self-avoiding
walks, as noted above.  In addition,
\begin{eqnarray}
&1=\mu_0^{(2)}<\mu_w^{(2)}<\mu_{w+1}^{(2)}<\mu_{2},
\label{e5} \\
&1 < \mu_{2} < \mu_w^{(3)}<\mu_{w+1}^{(3)}<\mu_{3}
\label{e6}
\end{eqnarray}
In addition, $\mu_2 = \mu_0^{(3)}$.

The number $c_n^{(w)}$ of self-avoiding walks from the origin in $S_w$, 
of length $n$, is a lower bound on the number of walks in $S_w(\lambda)$.  Thus,
$w_n(\lambda) \geq c_n^{(w)}$, since every walk in $S_w$ is also a walk in 
$\Sigma_w \supseteq S_w(\lambda)$.  Thus, 
$\liminf_{n\to\infty}\Sfrac{1}{n} \log w_n(\eps)
\geq \lim_{n\to\infty} \Sfrac{1}{n} \log c_n^{(w)}$.  If $\lambda\geq 1$, 
then $w_n(\lambda) = c_n^{(w)}$  with the result that
$\lim_{n\to\infty}\Sfrac{1}{n} \log w_n(\eps) 
= \lim_{n\to\infty} \Sfrac{1}{n} \log c_n^{(w)}$.  Using 
equation \Ref{e2} this gives theorem \ref{T2}.

\begin{theorem}
For all $\lambda \geq 0$,
\[ \liminf_{n\to\infty} \Sfrac{1}{n} \log w_n(\lambda)
\geq \lim_{n\to\infty} \Sfrac{1}{n} \log c_n^{(w)} = \log \mu_w^{(d)}.\]
If $\dps \lambda \geq 1$, $\rho^{(w)}(\lambda) 
=  \lim_{n\to\infty} \Sfrac{1}{n} \log w_n(\lambda) = \log \mu_w^{(d)}$. \qed
\label{T2}
\end{theorem}

\subsection{Ballistic walks in $S_w$}

\begin{figure}[h]
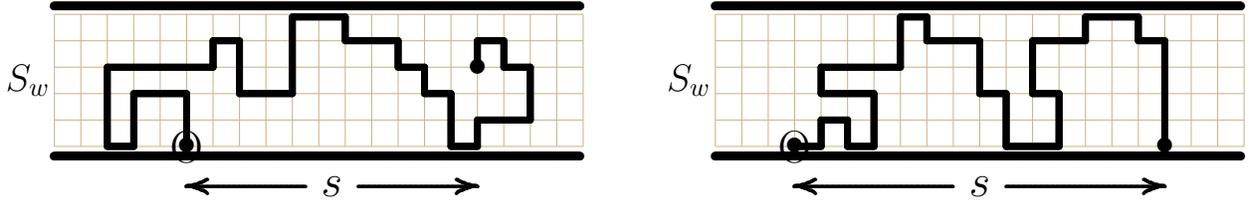


\beginpicture
\setcoordinatesystem units <1pt,1pt>
\setplotarea x from -160 to 300, y from -20 to 60
\multiput {\Large$S_w$} at -160 25 90 25 /

\setplotsymbol (.) \color{Tan}
\setplotarea x from -150 to 50, y from  0 to 50
\grid 20 5 
\setplotarea x from 100 to 300, y from 0 to 50
\grid 20 5
\setplotarea x from -160 to 300, y from -20 to 60

\setplotsymbol ({\footnotesize$\bullet$}) \color{Black}
\multiput {\Large$\bullet$} at -100 0 130 0 10 30 270 0 /
\plot -150 -4 50 -4 /
\plot -150 53 50 53 /
\plot 100 -4 300 -4 /
\plot 100 53 300 53 /

\multiput {\LARGE O} at -100 0 130 0 /

\setplotsymbol ({$\scalebox{2.5}{.}$})

\plot -100 0 -100 20 -120 20 -120 0 -130 0 -130 30 -90 30 -90 40 -80 40
-80 20 -60 20 -60 49 -40 49 -40 40 -20 40 -20 30 -10 30 -10 20 0 20 0 0 
10 0 10 10 30 10 30 30 20 30 20 40 10 40 10 30  /

\plot 130 0 140 0 140 10 150 10 150 0 160 0 160 20 140 20 140 30 
170 30 170 49 180 49 180 40 200 40 200 20 210 20 210 0 230 0 
230 20 220 20 220 40 240 40 240 49 260 49 260 40 270 40 270 0 /

\setplotsymbol ({$\scalebox{1.25}{.}$})
\arrow <8pt> [.2,.67] from -55 -15 to -100 -15 
\arrow <8pt> [.2,.67] from -35 -15 to 10 -15 
\put {\LARGE$s$} at -45 -15

\arrow <8pt> [.2,.67] from 190 -15 to 130 -15 
\arrow <8pt> [.2,.67] from 210 -15 to 270 -15 
\put {\LARGE$s$} at 200 -15

\color{black}\normalcolor
\endpicture

\caption{Ballistic walks $\omega=(\omega_0,\omega_1,\ldots,\omega_n)$
from the origin in $S_w$.
(Left) This walk has span $s=|x_1(\omega_0) - x_1(\omega_n)|$.  
(Right) An unfolded self-avoiding walk of span $s$.  In this walk 
$x_1(\omega_0)<x_1(\omega_i) \leq x_1(\omega_n)$ for all $1\leq i\leq n$
so that the origin is the unique left-most vertex, the first step from the origin is in 
the $x_1$-direction, and the last vertex $\omega_n$ is a right-most vertex.
Since the heights of the first and last vertices in $\omega$ are zero, this unfolded walk
is also a loop.}
\label{f4}
\end{figure}

A self-avoiding walk $\omega=(\omega_0,\omega_1,\ldots,\omega_n)$
of length $n$ in $S_w$ with $|\x1(\omega_0) - \x1(\omega_n)|=s$ is a 
\textit{ballistic walk} of \textit{span} $s$.  That is, the span of the ballistic walk is 
the absolute difference between the $x_1$-coordinates of its first and 
last vertices, and an example is illustrated in the left panel of figure \ref{f4}.

A ballistic walk $\omega$ of span $s$ is \textit{unfolded} if
$x_1(\omega_0)<x_1(\omega_i) \leq x_1(\omega_n)$.  
That is, the walk steps from its unique left-most vertex $\omega_0$  in the 
$x_1$-direction to $\omega_1$, and finally terminates in a right-most vertex 
$\omega_n$.  An unfolded walk is illustrated in the right panel in figure \ref{f4}.  
This walk is also a \textit{loop} of span $s$ (figure \ref{f4}(right)), which are
unfolded walks from the origin in $S_w$ with last vertex of 
\textit{height} $\x{d}(\omega_n)=0$.  Define
\begin{eqnarray*}
c_n^{(w)}(s) &=& \# \LC \hbox{ballistic walks from the origin in $S_w$ of length $n$ 
and span $s$} \RC\\
c_n^{(\dagger,w)}(s) &=& \# \LC \hbox{unfolded ballistic walks from the origin in 
$S_w$ of length $n$ and span $s$} \RC\\
\ell_n^{(w)}(s) &=& \# \LC \hbox{ballistic loops from the origin in $S_w$ of 
length $n$ and span $s$} \RC .
\end{eqnarray*}

\subsubsection{Ballistic loops in $S_w$}

A loop of length $n$ and span $s$ in $S_w$ can be concatenated with a loop of
length $m$ and span $t$ in $S_w$ by placing the first vertex of the second loop 
on the last vertex of the first loop.  The result is another loop of length $n{+}m$
and span $s{+}t$ in $S_w$.  Since there are $\ell_n^{(w)}(s)$ choices for the 
first loop, and $\ell_m^{(m)}(t)$ choices for the second loop, 
\begin{equation}
\ell_n^{(w)}(s)\,\ell_m^{(w)}(t) \leq \ell_{n+m}^{(w)}(s{+}t) .
\label{e7}
\end{equation}
Notice that $\ell_n^{(w)}(s)>0$ if $\lcl n/(w{+}1)\rcl \leq s\leq n$ in the square
lattice (the lower bound follows by packing a loop densely into a slit of height
$w$).  In three dimensions, $\ell_n^{(w)}(s)>0$ if $0 < s \leq n$.  Define
\begin{equation}
\lambda_0 = 
\cases{
1/(w{+}1), & \hbox{if $d=2$;} \cr
0, & \hbox{if $d=3$.} }
\end{equation}
Then the following theorem follows.

\begin{theorem}
The limit 
$\dps \log L_w(\lambda) 
= \lim_{n\to\infty} \Sfrac{1}{n} \log \ell_n^{(w)}(\lfl \lambda n\rfl)$
exists and is a concave function of $\lambda$ for $\lambda \in (\lambda_0,1)$.
Moreover, $\dps \sup_{\lambda\in(0,1)} \log L_w(\lambda) = \log \mu_w^{(d)}$.
\label{T3}
\end{theorem}

\proof
Existence of the limit and concavity follows from equation \Ref{e7} and by 
lemma 1 and theorem 2 in reference \cite{JvR21}. \qed
\bigskip

Since $\ell_n^{(w)}(n) = 2$ in both the square lattice and the cubic lattice it follows
that $\log L_w(1) = 0$.  It is also the case that $\log L_w(\lambda)$ is left-continuous 
at $\lambda=1$.  

\begin{lemma}
$\dps \lim_{\lambda\to 1^-} \log L_w(\lambda) = \log L_w(1) = 0$.  Thus, 
$\log L_w(\lambda)$ is left-continuous at $\lambda=1$, and therefore left-continuous 
on $(\lambda_0,1]$.
\label{L1}
\end{lemma}

\proof
Observe that $\ell_n^{(w)} (s) \geq 1$ for $\lceil n/(w+1) \rceil  \leq s \leq n$ in the
square lattice, and for $0<s\leq n$ in the cubic lattice.  Thus, for $\lambda \in (\lambda_0,1]$,
$\dps \log L_w(\lambda) 
= \lim_{n\to\infty} \Sfrac{1}{n} \log \ell_n^{(w)} (\lfl\lambda n\rfl) \geq 0$.

On the other hand, $\ell_n^{(w)}(s)$ is bounded above by the number of random walks
of length $n$ and with span $s$. Selecting $s$ steps of a random walk to
be East (to the right), and then over-counting by allowing the remaining $n{-}s$
steps to be in arbitrary directions,
\[ \ell_n^{(w)}(s) \leq \Bi{n}{s}\, (2d)^{n-s} . \]
Put $s=\lfl \lambda n \rfl$, take the power $1/n$ of the above, and let $n\to\infty$.
This gives
\[\log L_w(\lambda) 
\leq \lim_{n\to\infty} \log \LB \Bi{n}{\lfl \lambda n \rfl}^{1/n}\, (2d)^{1-\lfl \lambda n\rfl/n}\RB
= \log\LB \frac{(2d)^{1-\lambda}}{\lambda^\lambda\,(1-\lambda)^{1-\lambda}} \RB. \]
Take $\lambda\to 1^-$ to complete the proof. \qed
\bigskip

An additional and useful result is given by theorems 4 and  5 in reference \cite{JvR21}.

\begin{theorem}
Suppose that $(k_n)$ is a sequence such that $\lim_{n\to\infty} (k_n/n)=\lambda
\in (\lambda_0,1]$. Then the limit
\[ \lim_{n\to\infty} \Sfrac{1}{n} \log \ell_n^{(w)}(k_n) = \log L_w(\lambda) \]
exists. 
\qed
\label{T4}
\end{theorem}

Next, define the functions $\ell_n({\leq}s) = \sum_{t=0}^s \ell_n(t)$ and
$\ell_n({\geq}s) = \sum_{t=s}^n \ell_n(t)$.  By reference \cite{JvR21} the 
following limits exist.

\begin{theorem}
The following limits exist for $\lambda \in (\lambda_0,1)$
\[
\log L_w({\leq}\lambda) = \lim_{n\to\infty} \Sfrac{1}{n} \log \ell_n({\leq}\lfl \lambda n\rfl), 
\hspace{2mm}\hbox{and}\hspace{2mm}
\log L_w({\geq}\lambda) = \lim_{n\to\infty} \Sfrac{1}{n} \log \ell_n({\geq}\lfl \lambda n\rfl).
\]
$\log L_w({\leq}\lambda)$ and $\log L_w({\geq}\lambda)$ are concave functions 
on $(\lambda_0,1]$, $\log L_w(\lambda) 
= \min (\log L_w({\leq}\lambda),\log L_w({\geq}\lambda))$,
and $\max (\log L_w({\leq}\lambda),\log L_w({\geq}\lambda)) = \log \mu_w^{(d)}$.  
In addition, $\log L_w({\leq}\lambda)$ is non-decreasing while $\log L_w({\geq}\lambda)$ 
is non-increasing.
\label{T5}
\end{theorem}

\proof
The proof of this theorem follows from theorem 6 in reference \cite{JvR21}. \qed
\bigskip

The minimum span $s_n^{(min)}$ of a loop of length $n$ in $S_w$ for $w>0$ in the 
square lattice is at least $\lfl n / (w{+}1)\rfl$ and at most $\lcl n/(w{+}1)\rcl+1$. 
In the cubic lattice $s_n^{(min)}=1$ for all $n\geq 1$ and $w>0$.
\textit{Define}
\begin{equation}
\log L_w(\lambda_0) 
\stackrel{\hbox{def}}{=}  \lim_{\lambda\to\lambda_0^+} \log L_w(\lambda)
\geq \limsup_{n\to\infty} \Sfrac{1}{n} \log \ell_n^{(w)} (s_n^{(min)}) .
\label{e8}
\end{equation}
This defines $\log L_w(\lambda)$ to be a concave function on $[\lambda_0,1]$ in
the square and cubic lattices.  In the square lattice, $\lambda_0=1/(w{+}1)$
and $\log L_w(\lambda_0)=0$ if $w\in\{0,1\}$ and $\log L_w(\lambda_0)>0$ 
if $w\geq 2$. Notice that if $w=0$ then $\log L_w(\lambda_0)=0$ if $\lambda_0=1$
and it is not defined otherwise, and if $w=1$, then $1/2 \leq \lambda_0 \leq 1$.

In the cubic lattice $\log L_w(0) = \log \mu_w^{(2)}$ where 
$\mu_w^{(2)}$ is the square lattice growth contant of walks in a slit of width 
$w$. That is, $\log L_w(0)=0$ if $w=0$ and $\log L_w(0)>0$ if $w\geq 1$.  
Collecting these results gives
\begin{equation}
\log L_w(\lambda_0) \stackrel{\hbox{def}}{=} 
\lim_{\lambda\to\lambda_0^+} \log L_w(\lambda)
\cases{
\cases{
=0, & \hbox{if $w\leq 1$}; \\
>0, & \hbox{if $w\geq 2$};
}
& \hbox{if $d=2$}; \\
\cases{
=0, & \hbox{if $w=0$}; \\
>0, & \hbox{if $w\geq 1$};
}
& \hbox{if $d=3$}.
}
\label{e9}
\end{equation}

A pattern theorem \cite{HW62A,K63,K64} for walks and loops in $S_w$ in either the square or
cubic lattice was proven in reference \cite{SS98} for $w\geq 0$ (see section 
8.3 in reference \cite{JvR15}).  

\begin{lemma}
In the square lattice there exists a $\lambda_1>\lambda_0$ such that for all 
$\lambda \in [\lambda_0,\lambda_1)$,
\[ \log L_w(\lambda) = \lim_{n\to\infty} \Sfrac{1}{n} \log \ell_n^{(w)}(\lfl \lambda n\rfl) 
 < \lim_{n\to\infty} \Sfrac{1}{n} \log \ell_n^{(w)} = \log \mu_w^{(d)}  \]
provided that $w\geq 1$.
\label{L2}
\end{lemma}

\proof
Let $P$ be a pattern consisting of two consecutive
steps in the $x_1$-direction.  That is, $P=(\nu_0,\nu_1,\nu_2)$ with 
$x_1(\nu_0)+1=x_1(\nu_1)=x_1(\nu_2)-1$ and $x_j(\nu_0)=x_j(\nu_1)=x_j(\nu_2)$
for $2\leq j \leq d$.  In addition, let $P$ require all vertices $v\in S_w$ with 
$x_1(v)=x_1(\nu_1)$ to be unoccupied by the walk.  $P$ occurs at the $i$-th vertex
$\omega_i$ in a walk $\omega$ if $P$ can be translated such that $\nu_1=\omega_i$
and $\nu_j=\omega_{i+j-1}$ for $j=0,1,2$ while all vertices $v\in S_w$ with 
$x_1(v) = x_1(\omega_i)$ and $v\not=\omega_i$ is not in $\omega$ (that is, 
$v\not\in \omega$).

If $P$ occurs $k$ times in a walk $\omega$ of length $n$, then the span $s$ 
of $\omega$ is at least $k+\lfl n/(w{+}1)\rfl$.  Conversely, if the span of a walk is 
$s = k+\lfl n/(w{+}1)\rfl$, then the number of occurances of $P$ is at most $k$.  
Denote by $\ell_n^{(w)}$ the number of loops of length $n$ from the origin in $S_w$, 
and by $\ell_n^{(w)}(\#P{\leq}k)$ the number of loops from the origin 
in $S_w$ in which the pattern $P$ occurs \textit{at most} $k$ times.

Since $\ell_n^{(w)}(s)$ containes the pattern $P$ at most $s-\lfl n/(w{+}1)\rfl$ 
times, it follows that for $\lambda > 1/(w{+}1) = \lambda_0$,
\[ \ell_n^{(w)}(\lfl\lambda n \rfl) 
 \leq \ell_n^{(w)}(\#P{\leq}\lfl \lambda n \rfl{-}\lfl \lambda_0 n\rfl) \leq \ell_n^{(w)}. \]
Take the logarithms of these inequalities, divide by $n$ and take the limit superior
as $n\to\infty$.  This shows that
\[ \log L_w(\lambda) \leq \limsup_{n\to\infty} 
\Sfrac{1}{n} \log \ell_n^{(w)}(\#P{\leq}\lfl \lambda n \rfl{-}\lfl \lambda_0 n\rfl) 
\leq \log \mu_w^{(2)} . \]
By the pattern theorem for loops in $S_w$ there exists a $\lambda^*>\lambda_0$ such that
for all $\lambda \in [\lambda_0,\lambda^*]$,
\[  \limsup_{n\to\infty} \Sfrac{1}{n} 
\log \ell_n^{(w)}(\#P{\leq}\lfl \lambda n \rfl{-}\lfl \lambda_0 n\rfl) < \log \mu_w^{(2)} ,\]
in particular also since $\ell_n^{(w)}(\#P{\leq}k)$ is a non-decreasing function of $k$.
This shows that for all $\lambda \in [\lambda_0,\lambda^*]$, 
$\log L_w(\lambda) \leq \log \mu_w^{(2)}$. 
This completes the proof. \qed
\bigskip

Underlying lemma \ref{L2} is the fact that a square lattice self-avoiding walk 
in $S_w$ is ballistic.  That is, in the case of a loop of length $n$, the span $s = O(n)$.
Presumably this is not the case in $S_w$ in the cubic lattice where one expects $s=o(n)$.
Since loops of length $n$ and span $\lfl \lambda n \rfl$ are ballistic, it should be the
case that $\log L_w(\lambda)$ is a non-increasing function of $\lambda\in[\lambda_0,1]$
(and in the cubic lattice could be a strictly decreasing function of $\lambda\in [\lambda_0,1]$).  
This is stated as a conjecture.

\begin{conjecture}
In the cubic lattice $\log L_w(\lambda)$ (for $w\geq 0$) is a strictly decreasing function 
of $\lambda\in[\lambda_0,1]$. In addition, 
$$ 
\sup_{0<\lambda\leq 1} \log L_w(\lambda) 
= \lim_{\lambda\to 0^+} \log L_w(\lambda) 
= \log \mu_w^{(3)}. \eqno \qed
$$
\label{C1}
\end{conjecture}

By equation \Ref{e8} and theorem \ref{T3} the function $\log L_w(\lambda)$ is 
a concave function on $[\lambda_0,1]$.  By lemmas \ref{L1} and \ref{L2}
there exist $\lambda_1\in(\lambda_0,1)$ such that, in the square lattice, 
$\log L_w(\lambda) < \log L_w(\lambda_1) = \log \mu_w^{(d)}$.  In the cubic lattice,
$\lambda_0=0$ and $\lambda_1 \in [0,1)$, and similarly 
$\log L_w(\lambda_1) = \log \mu_w^{(d)}$.  By conjecture \ref{C1} it may be 
the case that $\lambda_1=0$ in the cubic lattice.

In other words, $\log L_w(\lambda)$ is strictly increasing in $[\lambda_0,\lambda_1)$
in the square lattice.  Since $\log L_w(\lambda)$ is concave on $[\lambda_0,1]$, it 
follows that it has a maximum at an $\lambda_2\in[\lambda_1,1]$ where 
$\lambda_2>\lambda_0$ in the square lattice if $w\geq 1$ and $\lambda_2\geq 0$
if $w\geq 0$ in the cubic lattice.

\begin{corollary}
For $w\geq 0$ in the square or cubic lattices, 
there exists critical values $\lambda_1,\lambda_2\in[\lambda_0,1)$ given by 
\[ \lambda_1 =\inf\{\lambda\svv \log L_w(\lambda) = \log \mu_w^{(d)}\}
\quad\hbox{and}\quad
\lambda_2 =\sup\{\lambda\svv \log L_w(\lambda) = \log \mu_w^{(d)}\}, \]
such that $\lambda_1\leq \lambda_2$ and in the square lattice $\lambda_0<\lambda_1$.
In addition, $\log L_w(\lambda) < \log L_w(\lambda_1) = \log \mu_w^{(d)}$ 
if $\lambda\in [\lambda_0,\lambda_1)$,
$\log L_w(\lambda) = \log \mu_w^{(d)}$ if $\lambda \in [\lambda_1,\lambda_2]$, and 
$\log L_w(\lambda) < \log L_w(\lambda_2) = \log \mu_w^{(d)}$ 
if $\lambda\in (\lambda_2,1]$. 
\label{cor1}
\end{corollary}

\proof
By theorem \ref{T3}, lemma \ref{L1} and equation \Ref{e8},
$\sup_{0<\lambda<1} \log L_w(\lambda) = \log \mu_w^{(d)}$ and 
$\log L_w(\lambda)$ is a concave function on $[\lambda_0,1]$. 

If $w=0$ in the square lattice, then trivially $\lambda_1=\lambda_2=1$.

If $w>0$ in the square lattice, or $w\geq 0$ in the cubic lattice, then 
define $\lambda_1$ and $\lambda_2$ as above.  Then $\lambda_0<\lambda_1\leq \lambda_2<1$
in the square lattice if $w>0$, and $0\leq\lambda_1\leq \lambda_2<1$ in the cubic lattice.
Thus, $\log L_w(\lambda) < \log L_w(\lambda_1) = \log \mu_w^{(d)}$ if 
$\lambda\in [\lambda_0,\lambda_1)$, by theorem \ref{T3} and lemma \ref{L2}.

By theorem \ref{T3} and lemma \ref{L1} it is necessarily the case that 
$\lambda_2<1$ if $w>0$ in the square lattice, or $w\geq 0$ in the cubic lattice.
By the definition of $\lambda_2$, 
$\log L_w(\lambda) < \log L_w(\lambda_2) = \log \mu_w^{(d)}$ 
if $\lambda\in (\lambda_2,1]$.\qed
\bigskip

In view of corollary \ref{cor1} the following conjectures:

\begin{conjecture}
In corollary \ref{cor1}, $\lambda_1=\lambda_2$. \qed
\label{C2}
\end{conjecture}

This is consistent with conjecture \ref{C3}.

\begin{conjecture}
For $w>0$ in the square lattice, or $w\geq 0$ in the cubic lattice, the
function $\log L_w(\lambda)$ is a strictly concave function on $[\lambda_0,1]$. \qed
\label{C3}
\end{conjecture}

By theorem \ref{T2} and corollary \ref{cor1}, $\log L_w({\leq}\lambda)$ is 
strictly increasing for $\lambda\in(\lambda_0,\lambda_1)$ if $\lambda_0<\lambda_1$.
It follows that $\lim_{\lambda\to \lambda_0^+} \log L_w({\leq}\lambda) 
= \lim_{\lambda\to \lambda_0^+} \log L_w(\lambda) = \log L_w(\lambda_0^+)$ 
in both the square lattice (if $w>0$) and in the cubic lattice
(if $w\geq 0$).  

Defining $\log L_w({\leq}\lambda_0) = \log L_w(\lambda_0)$ and
$\log L_w({\leq}1) = \log \mu_w^{(d)}$, and similarly for $\log L_w({\geq}\lambda)$, 
the domains of these functions are extended to $[\lambda_0,1]$.  Moreover, 
by theorems \ref{T3} and \ref{T2}, and since 
$\ell_n^{(w)}({\leq}s) + \ell_n^{(w)}({\geq}s) \geq \ell_n^{(w)}$, 
it follows that for each $\lambda\in[\lambda_0,1]$,
\begin{equation}
\max_{\lambda\in[\lambda_0,1]} 
(\log L_w({\leq}\lambda),\log L_w({\geq}\lambda)) = \log \mu_w^{(d)}.
\label{e10}
\end{equation}

\begin{figure}[h]
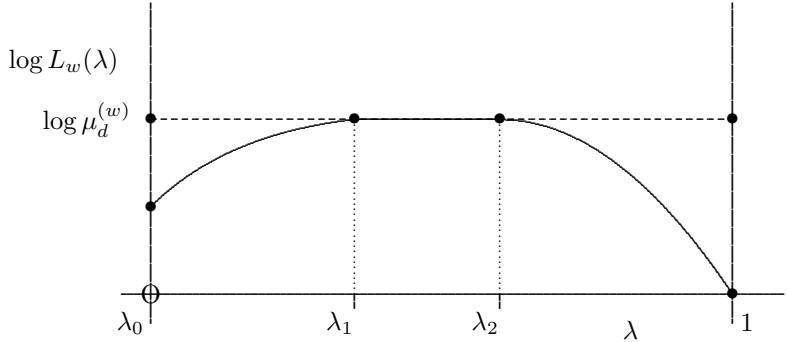

\beginpicture
\setcoordinatesystem units <1.1pt,1.1pt>
\setplotarea x from -100 to 300, y from -20 to 100

\plot -10 0 220 0 /  \plot 0 -10 0 100 / \put {O} at 0 0
\plot 200 -10 200 100 / \put {1} at 205 -10
\put {$\log L_w(\lambda)$} at -30 80 
\put {$\log \mu_d^{(w)}$} at -22 60 
\put {$\lambda$} at 165 -12.5

\plot 70 -5 70 0 /  \plot 120 -5 120 0 /
\put {$\lambda_1$} at 65 -10
\put {$\lambda_2$} at 115 -10 
\put {$\lambda_0$} at -7 -10 

\multiput {$\bullet$} at 0 30 0 60 200 60 70 60 120 60 200 0  /

\setdashes <2pt>
\plot 0 60 200 60 /

\setdots <2pt>
\plot 70 0 70 60 / \plot 120 0 120 60 /

\setsolid
\plot 70 60 120 60 /
\setquadratic
\plot 0 30 30 50 70 60 / 
\plot 120 60 160 45 200 0 /

\endpicture
\caption{A schematic plot of $\log L_w(\lambda)$.  The function is 
concave and has a maximum value $\log \mu_d^{(w)}$ (theorem \ref{T3}).
As $\lambda\to 1^-$ it approaches zero (lemma \ref{L1}).  In the square lattice
$\lambda_0 = 1/(w{+}1)$ and in the cubic lattice $\lambda_0=0$.  By corollary \ref{cor1}
there exists $\lambda_1\leq \lambda_2 < 1$ such that 
$\log L_w(\lambda) = \log \mu_w^{(d)}$ when $\lambda \in [\lambda_1,\lambda_2]$.
In the square lattice $\lambda_0<\lambda_1$ (by lemma \ref{L2}).  By corollary \ref{cor2}
$\log L_w({\leq}\lambda) = \log L_w(\lambda)$ if $\lambda \in [\lambda_0,\lambda_2]$
while $\log L_w({\leq}\lambda) = \log \mu_w^{(d)}$ if $\lambda \in [\lambda_1,1]$.
Similarly, $\log L_w({\geq}\lambda) = \log L_w(\lambda)$ if $\lambda \in [\lambda_1,1]$
while $\log L_w({\geq}\lambda) = \log \mu_w^{(d)}$ if $\lambda \in [\lambda_0,\lambda_2]$.
By conjectures \ref{C1} and \ref{C2} it may be the case that 
$\lambda_2 = \lambda_1=\lambda_0 = 0$ in the cubic lattice.  By conjecture \ref{C3} 
the graph may be strictly concave, in which case $\lambda_1=\lambda_2$ and it 
has a maximum at exactly one point. }
\label{F5}
\end{figure}

\begin{corollary}
If $w>0$ in the square lattice, or $w\geq 0$ in the cubic lattice, 
there exists a critical value $\lambda_2\in[\lambda_0,1)$, such that for each 
$\lambda\in (\lambda_2,1]$, $\log L_w({\leq}\lambda) = \log \mu_w^{(d)}$ and 
$\log L_w(\lambda) = \log L_w({\geq}\lambda) < \log \mu_w^{(d)}$.
\label{cor2}
\end{corollary}

\proof
Since  $\log L_w(\lambda) = \min (\log L_w({\leq}\lambda), \log L_w({\geq}\lambda))$ 
and $\log L_w({\leq}\lambda)$ is non-decreasing while $\log L_w({\geq}\lambda)$ is 
non-increasing (and these functions are continuous on $[\lambda_0,1]$), it follows by 
theorem \ref{T3} that
\[ \lim_{\lambda\to 1^-} \log L_w({\geq}\lambda) = 0 .\]
By corollary \Ref{cor1} there exists an $\lambda_2\in[\lambda_0,1)$ such that for
each $\lambda\in [\lambda_2,1]$, $\log L_w({\leq}\lambda) = \log \mu_w$ and
$L_w({\geq}\lambda) < \log \mu_w$. \qed
\bigskip

\begin{corollary}
If $w\geq 1$ in the square lattice, or $w\geq 0$ in the cubic lattice, then
for $\lambda\in [0,\lambda_1]$, $\log L_w(\lambda) = \log L_w({\leq}\lambda)$.
Similarly, for $\lambda\in [\lambda_2,1]$, 
$\log L_w(\lambda) = \log L_w({\geq}\lambda)$. 
In addition, if $\lambda\in[\lambda_1,\lambda_2]$, then 
$\log L_w(\lambda) = \log L_w({\leq}\lambda) = \log L_w({\geq}\lambda) 
= \log \mu_w^{(d)}$.  Moreover $\dps \lim_{\lambda\to 1^-} \log L(\lambda) 
= \lim_{\lambda\to 1^-} \log L({\geq}\lambda) = 0$ so that $\lambda_1\leq \lambda_2 < 1$. \qed
\label{cor3}
\end{corollary}

\subsubsection{Unfolded ballistic walks in a slit or a slab:}

Denote the number of unfolded walks from the origin, of length $n$ in $S_w$, 
with span $s$, by $c_n^{(\dagger,w)}(s)$.  Clearly, 
$\ell_n^{(w)}(s) \leq c_n^{(\dagger,w)}(s)$,  since each loop is also an unfolded walk.  
This shows that for $\lambda\in[\lambda_0,1]$,
\begin{equation}
\log L_w(\lambda) \leq \liminf_{n\to\infty} \Sfrac{1}{n} 
\log c_n^{(\dagger,w)}(\lfl \lambda n \rfl).
\label{e11}
\end{equation}

\begin{figure}[h]
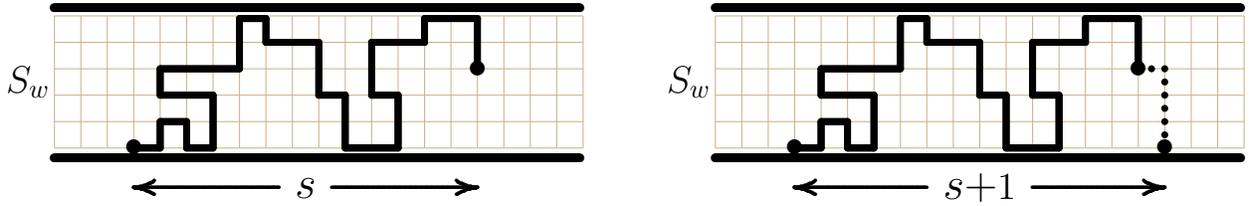

\beginpicture
\setcoordinatesystem units <1pt,1pt>
\setplotarea x from -160 to 300, y from -20 to 60
\multiput {\Large$S_w$} at -160 25 90 25 /

\setplotsymbol (.) \color{Tan}
\setplotarea x from -150 to 50, y from  0 to 50
\grid 20 5 
\setplotarea x from 100 to 300, y from 0 to 50
\grid 20 5
\setplotarea x from -160 to 300, y from -20 to 60

\setplotsymbol ({\footnotesize$\bullet$}) \color{Black}
\multiput {\Large$\bullet$} at -120 0 130 0 10 30 270 0 260 30  /
\plot -150 -4 50 -4 /
\plot -150 53 50 53 /
\plot 100 -4 300 -4 /
\plot 100 53 300 53 /

\setplotsymbol ({$\scalebox{2.5}{.}$})

\plot -120 0 -110 0 -110 10 -100 10 -100 0 -90 0 -90 20 -110 20 -110 30 
-80 30 -80 49 -70 49 -70 40 -50 40 -50 20 -40 20 -40 0 -20 0 
-20 20 -30 20 -30 40 -10 40 -10 49 10 49 10 30 /

\plot 130 0 140 0 140 10 150 10 150 0 160 0 160 20 140 20 140 30 
170 30 170 49 180 49 180 40 200 40 200 20 210 20 210 0 230 0 
230 20 220 20 220 40 240 40 240 49 260 49 260 30 /

\setdots <5pt>
\plot 260 30 270 30 270 0 /
\setsolid

\setplotsymbol ({$\scalebox{1.25}{.}$})
\arrow <8pt> [.2,.67] from -65 -15 to -120 -15 
\arrow <8pt> [.2,.67] from -45 -15 to 10 -15 
\put {\LARGE$s$} at -55 -15

\arrow <8pt> [.2,.67] from 180 -15 to 130 -15 
\arrow <8pt> [.2,.67] from 220 -15 to 270 -15 
\put {\LARGE$s{+}\hbox{\Large$1$}$} at 200 -15

\color{black} \normalcolor
\endpicture
\caption{An unfolded walk (left) of span $s$ in a slit $S_w$ of width $w$.
This walk can be turned into a loop (right) of span $s{+}1$ by 
appending at most $w{+}1$ steps on the right to reattach the endpoint 
to the bottom boundary of $S_w$.}
\label{F6}
\end{figure}

On the other hand, each unfolded walk of length $n$ can be turned into an 
unfolded loop by adding no more than $w{+}1$ steps at the end of the walk 
(see figure \ref{F6}) to reconnect its endpoint with the bottom boundary of 
$S_w$.  This shows that
\begin{equation}
c_n^{(\dagger,w)}(s) 
\leq \sum_{j=1}^{w+1} \ell_{n+j}^{(w)}(s{+}1) .
\label{e12}
\end{equation}
This gives the following lemma.

\begin{lemma}
If $\lambda \in [\lambda_0,1]$, then the limit 
$\dps \lim_{n\to\infty} \Sfrac{1}{n} \log c_n^{(\dagger,w)}(\lfl \lambda n \rfl)
= \log L_w(\lambda)$ exists. 
\label{L3}
\end{lemma}

\proof
Add $w{+}1{-}j$ horisontal steps to each loop counted by the right hand side
of equation \Ref{e12}.  Then all the loops have length $n{+}w{+}1$, and
spans in $\{s{+}1,s{+}2,\ldots,s{+}w{+}2\}$.  That is
$$ c_n^{(\dagger,w)}(s) 
\leq \sum_{j=1}^{w+1} \ell_{n+w+1}^{(w)}(s{+}w{+}2{-}j) . \eqno (*) $$
By theorem \ref{T4}, for each fixed $j$,
$\dps \lim_{n\to\infty} \Sfrac{1}{n} 
 \log \ell_{n+w+1}^{(w)}(\lfl \lambda n\rfl{+}w{+}2{-}j) = \log L_w(\lambda)$.
Thus, taking logs of equation (*) and putting $s=\lfl \lambda n\rfl$, 
dividing by $n$, and then letting $n\to\infty$,
\[ \limsup_{n\to\infty} \Sfrac{1}{n} \log c_n^{(\dagger,w)}(\lfl \lambda n \rfl)
\leq \lim_{n\to\infty} \Sfrac{1}{n} \log
 \sum_{j=1}^{w+1} \ell_{n+w+1}^{(w)}(\lfl \lambda n\rfl {+}w{+}2{-}j) 
= \log L_w(\lambda) . \]
By equation \Ref{e11} this proves existence of the limit.  This completes
the proof. \qed
\bigskip

Consider next $c_n^{\dagger,w}({\leq}t)$ and $c_n^{\dagger,w}({\geq}t)$
defined by
\begin{equation}
c_n^{(\dagger,w)}({\leq}t) = \sum_{s=0}^t c_n^{(\dagger,w)}(s),
\quad\hbox{and}\quad
   c_n^{(\dagger,w)}({\geq}t) = \sum_{s=t}^n c_n^{(\dagger,w)}(s). 
   \end{equation}
   
\begin{lemma}
The following limits exist for $\lambda\in[\lambda_0,1]$:
\[
\lim_{n\to\infty} \Sfrac{1}{n} \log c_n^{(\dagger,w)} ({\leq}\lfl \lambda n\rfl)
= \log L_w({\leq}\lambda) , 
\hspace{2mm}\hbox{and}\hspace{2mm}
\lim_{n\to\infty} \Sfrac{1}{n} \log c_n^{(\dagger,w)} ({\geq}\lfl \lambda n\rfl)
= \log L_w({\geq}\lambda) .
\]
\end{lemma}

\proof
By appending $w{+}1{-}j$ edges to the endpoint of the loops on the 
right hand side of equation \Ref{e12},
\[ c_n^{(\dagger,w)}({\leq}s) \leq \sum_{j=1}^{w+1} \ell_{n+j}^{(w)}({\leq}s{+}1)
   \leq \sum_{j=1}^{w+1} \ell_{n+w+1}^{(w)}({\leq}s{+}w{+}2{-}j)
   \leq (w{+}1)\, \ell_{n+w+1}^{(w)}({\leq}s{+}w{+}2)\]
Consider the term on the right hand side above if $s=\lfl\lambda n\rfl$.  This is
\[ \ell_{n+w+1}^{(w)}({\leq}\lfl \lambda n\rfl{+}w{+}2) 
= \sum_{k=0}^{\lfl \lambda n \rfl{+}w{+}2} \ell_{n+w+1}^{(w)}(k)
\leq (\lfl \lambda n \rfl{+}w{+}2)\, \ell_{n+w+1}^{(w)}(\gamma_n)\]
where $k=\gamma_n \in [0,\lfl \eps n \rfl{+}w{+}2]$ maximizes the 
terms in the summation for each value of $n$.  Take logarithms, divide by
$n$, and then take $n\to\infty$. The sequence of points $(\gamma_n/(n{+}w{+}1)$
has an accumulation point $\gamma\in[0,\lambda]$ and by theorem \ref{T4},
this gives
\begin{eqnarray*}
& \limsup_{n\to\infty} \Sfrac{1}{n} 
\log  c_n^{(\dagger,w)}({\leq}\lfl \lambda n\rfl{+}w{+}2) \cr
& \hspace{2cm} \leq \limsup_{n\to\infty} \Sfrac{1}{n} 
\log  \ell_{n+w+1}^{(w)}({\leq}\lfl \lambda n\rfl{+}w{+}2) 
\leq \log L_w(\gamma)\leq \log L_w({\leq}\lambda) 
\end{eqnarray*}
since $\gamma \leq \lambda$.

On the other hand, $\ell_n^{(w)}({\leq}s) \leq c_n^{(\dagger,w)}({\leq}s)$, so that
\[\log L_w({\leq} \lambda) 
= \lim_{n\to\infty} \Sfrac{1}{n} \log \ell_n^{(w)}({\leq}\lfl\lambda n\rfl) 
\leq \liminf_{n\to\infty} \Sfrac{1}{n} \log c_n^{(\dagger,w)}({\leq}\lfl\lambda n\rfl).\]
This completes the proof for $c_n^{(\dagger,w)}({\leq}\lfl\lambda n\rfl)$.

The proof for $c_n^{(\dagger,w)}({\geq}\lfl\lambda n\rfl)$ is similar. \qed
\bigskip

\subsubsection{Ballistic walks in slits and slabs:}
Now consider a self-avoiding walk in $S_w$, with its span (the distance between 
its endpoints) in the $x_1$-direction equal to $s$ (as illustrated in the left panel 
in figure \ref{f4}).  Let the number of such walks, from the origin, of length $n$ 
with span between endpoints equal to $s$, be $c_n^{(w)}(s)$.  

If $d=2$ and $w=0$, then $c_0^{(0)}(0) = 1$ and $c_n^{(0)}(n) = 2$ if
$n>0$, otherwise it is zero.  This shows that 
\begin{equation}
\lim_{n\to\infty} \Sfrac{1}{n} \log c_n^{(0)}(\lfl \lambda n \rfl) = 0
\quad\hbox{if $\lambda=1$},
\label{e14}
\end{equation}
and otherwise this limit is equal to $-\infty$.

Thus, assume that $w\geq 1$ if $d=2$.

By unfolding walks of length $n$ \cite{HW62A} and of span $s=\lfl \lambda n\rfl$ 
in the $x_1$-direction one obtains
\begin{equation}
\ell_n^{(w)}(\lfl \lambda n \rfl) \leq c_n^{(w)}(\lfl \lambda n \rfl)
\leq e^{o(n)} \sum_{s=\lfl \lambda n \rfl}^n c_n^{(\dagger,w)}(s)
= e^{o(n)}\, c_n^{(\dagger,w)}({\geq}\lfl \lambda n \rfl) ,
\end{equation}
since unfolding a walk will increase its span.  Applying the construction in 
figure \ref{F5} to the right hand side of this equation,
\begin{equation}
\ell_n^{(w)}(\lfl \lambda n \rfl) \leq c_n^{(w)}(\lfl \lambda n \rfl)
\leq (w{+}1)\,e^{o(n)}\, \ell_{n+w+1}^{(w)}({\geq}\lfl \lambda n \rfl) .
\end{equation}
Taking logarithms, dividing by $n$ and letting $n\to\infty$ give, by
theorems \ref{T3} and \ref{T5},
\begin{equation}
\log L_w(\lambda) 
\leq \liminf_{n\to\infty} \Sfrac{1}{n} \log c_n^{(w)}(\lfl \lambda n \rfl)
\leq \limsup_{n\to\infty} \Sfrac{1}{n} \log c_n^{(w)}(\lfl \lambda n \rfl)
\leq \log L_w({\geq}\lambda) .
\end{equation}
By corollary \ref{cor1}, there exists $\lambda_1,\lambda_2 \in [\lambda_0,1)$
such that $\lambda_1 \leq \lambda_2$ and if $\lambda\in[\lambda_1,1]$ then 
$\log L_w(\lambda)  = \log L_w({\geq}\lambda)$.  Putting $\lambda_0=\lambda_1=1$
when $d=2$ and $w=0$ this gives, with equation \Ref{e14}, theorem \ref{T6}.

\begin{theorem}
Let $\lambda\in [\lambda_1,1]$.  Then the limit
$\dps \log C_w(\lambda) =
\lim_{n\to\infty} \Sfrac{1}{n} 
\log c_n^{(w)}(\lfl \lambda n \rfl) = \log L_w(\lambda)$ exists. \qed
\label{T6}
\end{theorem}

In the event that $\lambda\in[\lambda_0,\lambda_1)$ \textit{define}
\begin{equation}
\dps \log C_w(\lambda) =
\limsup_{n\to\infty} \Sfrac{1}{n} \log c_n^{(w)}(\lfl \lambda n \rfl) 
\end{equation}

The pattern theorem for walks and loops in $S_w$ \cite{SS98} for $w\geq 1$
shows that, in the square lattice, 
\begin{equation}
\log C_w(\lambda) 
= \limsup_{n\to\infty} \Sfrac{1}{n} \log c_n^{(w)}(\lfl \lambda n \rfl) 
< \log \mu_w^{(2)}, \quad\hbox{if $|\lambda-\lambda_0|$ is small.}
\end{equation}
The proof of this is similar to the proof of lemma \ref{L2}.   In view of 
this a corollary of theorem \ref{T6} and corollary \ref{cor3} is the following.

\begin{corollary}
If $w\geq 1$ in the square lattice then there exists a $\lambda_1^\prime$ such
that $\lambda_0 < \lambda_1^\prime \leq \lambda_1 \leq \lambda_2 < 1$.
In addition
\[ \log C_w(\lambda)  
\cases{
< \log \mu_w^{(2)}, & \hbox{if $\lambda\in[\lambda_0,\lambda_1^\prime)$}; \\
= \log \mu_w^{(2)}, & \hbox{if $\lambda\in[\lambda_1^\prime,\lambda_2]$};\\
< \log \mu_w^{(2)}, & \hbox{if $\lambda\in(\lambda_2,1)$}; \\
= 0, & \hbox{if $\lambda=1$}.
} \]
\label{cor4}
\end{corollary}

\proof
The proof proceeds by showing that there exists a $\lambda^\prime_1\in
(\lambda_0,\lambda_1]$ such that $\log C_w(\lambda) < \log \mu_w^{(2)}$
if $\lambda \in [0,\lambda_1^\prime]$.

Since $\ell_w^{(2)}(\lfl \lambda n \rfl) \leq c_n^{(w)}(\lfl \lambda n \rfl)$ it
follows that $\log L_w(\lambda) \leq C_w(\lambda)$. Since $\log L_w(\lambda) 
= \log \mu_w^{(2)}$ in the square lattice for $\lambda\in[\lambda_1,\lambda_2]$ 
and $\log C_w(\lambda) < \log \mu_w^{(2)}$ for $\lambda-\lambda_0$ 
small positive, there exists a $\lambda_1^\prime \leq \lambda_1$ as claimed.

The remaining relations are consequences of corollaries \ref{cor1} and \ref{cor3}. \qed
\bigskip

We conjecture that $\log C_w(\lambda)$ exists as a limit for all $w\geq 0$, and 
that $\log C_w(\lambda) = \log L_w(\lambda)$ on $[\lambda_0,1]$ in the square
lattice.

In the cubic lattice $\log C_w(\lambda)$ is bounded by the following corollary.

\begin{corollary}
In the cubic lattice $\lambda_0=0 \leq \lambda_1 \leq \lambda_2 < 1$.  Then
\[ \log C_w(\lambda)  
\cases{
= \log \mu_w^{(3)}, & \hbox{if $\lambda\in[\lambda_1,\lambda_2]$};\\
< \log \mu_w^{(3)}, & \hbox{if $\lambda\in(\lambda_2,1)$}; \\
= 0, & \hbox{if $\lambda=1$}.
} \]
\label{cor5}
\end{corollary}

\proof
This follows from corollaries \ref{cor1} and \ref{cor3}. \qed
\bigskip

In the cubic lattice, we conjecture that $\lambda_1=\lambda_2=0$ (that is, 
$\log C_w(0) = \log \mu_w^{(3)}$ and $\log C_w(\lambda) < \log \mu_w^{(3)}$
if $\lambda \in (0,1]$.

Finally, in view of conjecture \ref{C1} the following for ballistic cubic lattice 
walks in $S_w$.  A walk $\omega=\{\omega_0,\omega_1,\ldots,\omega_n\}$ 
with $\omega_0$ at the origin and $\omega_n=(0,1,0)$ has span $0$ and may 
be turned into a polygon by adding an edge to join its endpoints.  This gives a 
polygon rooted at the origin and the vertex $(0,1,0)$ of length $n{+}1$ 
in $L_w$.  Conversely, a polygon rooted in the edge joining the origin to
$(0,1,0)$ can be turned into a self-avoiding walk of span $0$.  Denote the 
number of such rooted polygons of length $n{+}1$ in $S_w$ by $p_{n{+}1}^{(w)}$.  
It follows that
\begin{equation}
c_n^{(w)}(0) \geq p_{n+1}^{(w)} .
\end{equation}
By reference \cite{SSW88A} it follows that 
\begin{equation}
\log \mu_w^{(3)} = \lim_{n\to\infty} \Sfrac{1}{n{+}1} \log p_{n+1}
\leq \liminf_{n\to\infty} \Sfrac{1}{n} \log c_n^{(w)}(0) 
\leq \limsup_{n\to\infty} \Sfrac{1}{n} \log c_n^{(w)}(0) \leq \log \mu_w^{(3)}.
\end{equation}
Since $\lambda_0 = 0$ in the cubic lattice, this shows that 
\begin{equation}
\lim_{n\to\infty} \Sfrac{1}{n} \log c_n^{(w)}(\lfl\lambda_0n\rfl)
= \log C_w(\lambda_0)
= \log \mu_w^{(3)}.
\end{equation}
By corollary \ref{cor5} this shows that
$\log C_w(\lambda_0) = \log C_w(\lambda) = \log \mu_w^{(3)}$
for $\lambda \in [\lambda_1,\lambda_2]$.  It is not known that
$\log C_w(\lambda)$ is monotone for $\lambda \in [0,1]$.  However,
the above is consistent with $\lambda_0=\lambda_1=\lambda_2=0$ in which
case $\log C_w(\lambda)$ exists and is strictly decreasing in $[0,1]$.

\begin{figure}[t]
\centering
\includegraphics[width=0.9\textwidth]{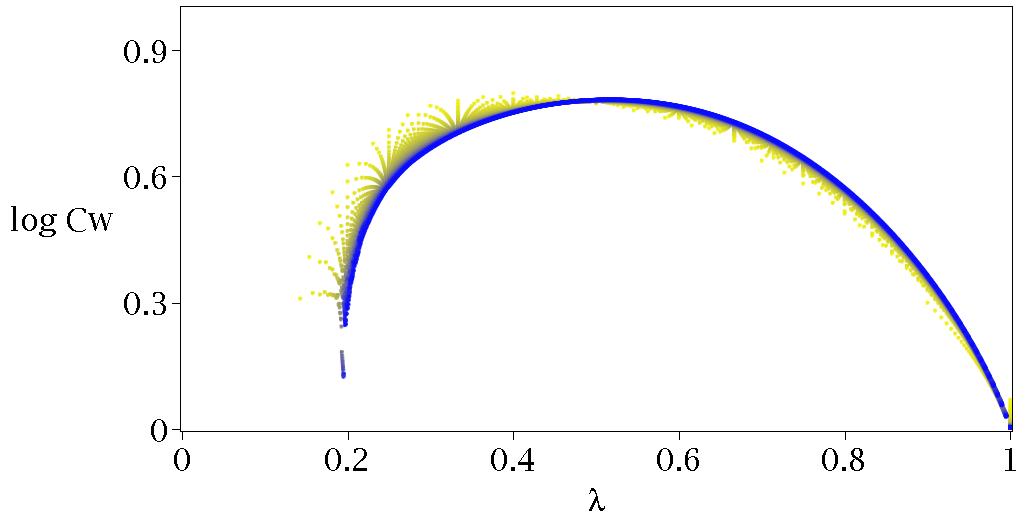}
\caption{Finite size approximations to $\log C_w(\lambda)$ for $10\leq n\leq200$
in the square lattice and for $w=4$.  Larger values of $n$ correspond to darker 
points in the plot.  As $n$ increases, the points accumulate close to the limiting curve 
which is a concave curve approaching $0$ as $\lambda \to 1^-$.  The maximum in the
data corresponding to $n=200$ is approximately located at $\lambda \approx 0.526$.}
\label{f7}
\end{figure}

\begin{figure}[t]
\centering
\includegraphics[width=0.9\textwidth]{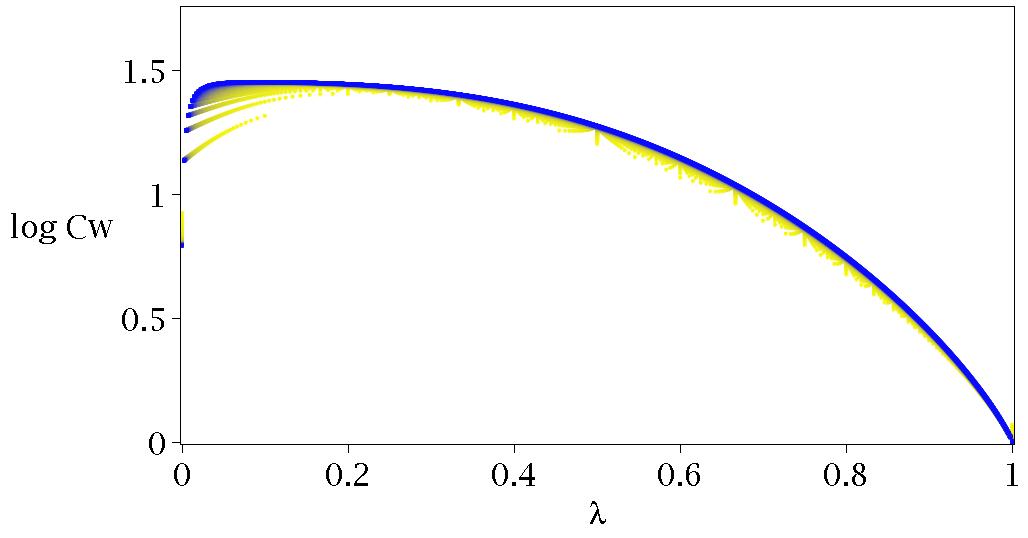}
\caption{Finite size approximations to $\log C_w(\lambda)$ for $10\leq n\leq400$
in the cubic lattice and for $w=4$.  Larger values of $n$ correspond to darker 
points in the plot.  As $n$ increases, the points accumulate close to the limiting curve 
which is a concave curve approaching $0$ and $\lambda \to 1^-$.  The maximum in the
data corresponding to $n=400$ is approximately located at $\lambda \approx 0.05$.}
\label{f8}
\end{figure}

\subsubsection{Numerical results:}
\label{S213}

Numerical estimates of $\log C_w(\lambda)$ can be obtained by sampling 
self-avoiding walks in slits and slabs.  This was done using the PERM algorithm 
\cite{G97}, in its flatPERM \cite{PK04} and parallel implementation \cite{CJvR20}, 
to sample self-avoiding walks from the origin in a slit or slab to length $200$ in the 
square lattice, and $400$ in the cubic lattice, while keeping track of the span 
of the walk (the \textit{span} as defined in figure \ref{f4}).  Data were collected 
for slits of width $1\leq w \leq 5$ in the square lattice and $0\leq w \leq 5$ in 
the cubic lattice.  In particular, the algorithm gives numerical approximations 
to $c_n^{(w)}(s)$,  namely the number of walks from the origin in $S_w$ 
and of length $n$ and span $s$.   The details of the calculations are shown 
in table \ref{tafel1}. In figure \ref{f7} we plot the square lattice finite 
approximation 
\begin{equation}
\log C_w(\lambda) \approx \frac{1}{n} \log c_n^{(w)}(\lfl \lambda n \rfl)
\end{equation}
determined from our data for $10\leq n \leq 200$ for $w=4$.  The curve
appears to have a single maximum, which is an estimate of $\lambda_2$
in figure \ref{F5}. It is also consistent with 
$\lambda_1^\prime = \lambda_1=\lambda_2$.  The cubic lattice 
approximations are similarly plotted in figure \ref{f8}.  In this case, the data
appear to converge to a non-increasing limiting curve consistent with 
$\lambda_0=\lambda_1=\lambda_2=2$. Finite size estimates of $\lambda_2$
are listed in table \ref{tafel1}.

\begin{table}[h]
\caption{Numerical estimates of $\mu_w^{(d)}$ and $\lambda_2$}
\begin{indented}
\lineup
\item[]
\begin{tabular}{llllll}
\br                              
$w$ &  & Iterations & $n_{max}$ & $\mu_w^{(d)}$ & $\lambda_2$   \cr 
\mr
$0$ & \multirow{6}{*}{square lattice} & -- & -- & $1$ (exact) & -- \cr
$1$ &       & $4\times 10^6$ & $200$ & $(1{+}\sqrt{5})/2^{\;*}$ & $0.723(2)$ \cr
$2$ &       & $4\times 10^6$ & $200$ & $1.9144(3)$ & $0.624(1)$ \cr
$3$ &       & $4\times 10^6$ & $200$ & $2.0873(1)$   & $0.567(2)$ \cr
$4$ &       & $8\times 10^6$ & $200$ & $2.1990(2)$ & $0.526(2)$ \cr
$5$ &       & $1\times 10^7$ & $200$ & $2.2770(9)$ & $0.499(2) $ \cr
\mr
$0$ & \multirow{6}{*}{cubic lattice} & $5\times 10^6$ & $400$ & $2.63815(2)$ $\dagger$ & $0.07(5)$ \cr
$1$ &       & $6\times 10^6$ & $400$ & $3.55333(4) $ & $0.06(4)$  \cr
$2$ &       & $8\times 10^6$ & $400$ & $3.9456(2)$ & $0.05(3)$ \cr
$3$ &       & $1\times 10^7$ & $400$ & $4.1577(4)$ & $0.04(3)$ \cr    
$4$ &       & $1\times 10^7$ & $400$ & $4.2869(4)$ & $0.05(4)$ \cr
$5$ &       & $1\times 10^7$ & $400$ & $4.3719(4)$ & $0.04(3)$ \cr
\br
\end{tabular}
\begin{tablenotes}
\item * Sloane A038577, \cite{N12}, Numerical estimate: $1.6185(4)$ 
\item $\dagger$ $\mu_2 = 2.63815853032790(3)$ \cite{JSG16}
\end{tablenotes}
\end{indented}
\label{tafel1}   
\end{table}

In table \ref{tafel1} the details and results of the simulations are shown.  
Estimates for $\mu_w^{(d)}$ are shown, as well as estimates for $\lambda_2$
(the location of the maximum in $\log C_w(\lambda)$).  If $d=2$ and $w=0$, 
then $\mu_0^{(2)}=1$.  If $w=1$ instead, then $\mu_1^{(2)}=2/(\sqrt{5}-1)$ 
(Sloane online sequences of integers A038577; see reference \cite{N12}).  

\begin{figure}[t]
\centering
\includegraphics[width=0.9\textwidth]{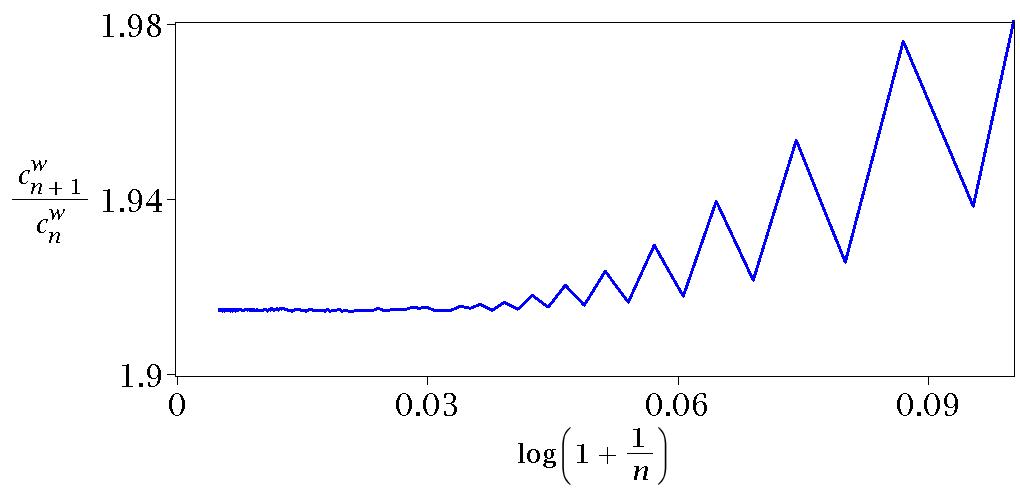}
\caption{Finite size approximations to $\mu_w^{(d)}$ for $10\leq n\leq200$
in the square lattice and for $w=2$.  The initial partiy effects die down as
$n$ increases, and the estimate stabilize well towards the $y$-axis.  Extrapolating
this gives an estimate of $\mu_w^{(d)}$.}
\label{f9}
\end{figure}

In order to estimate $\mu_w^{(w)}$ for larger values of $w$, assume that
\begin{equation}
c_n^{(w)} = A_w\, n^{\gamma_w-1}\,(\mu_w^{(d)})^n \,(1+o(1)).
\end{equation}
Expanding and simplifying the log of $c_{n+1}^{(w)}/c_n^{(w)}$ give
\begin{equation}
\log (c_{n+1}^{(w)} / c_n^{(w)})
= (\gamma_w-1) \log (1{+}1/n) + \log \mu_w^{(d)} + o(1) .
\end{equation}
Plotting the left hand side as a function of $1/n \approx \log (1+1/n)$ gave graphs 
such as in figure \ref{f9} (for $w=2$).  Fitting a quadratic curve to the data points
with $n\geq 40$ gives our best estimate of $\mu_w^{(d)}$.  This is shown in 
table \ref{tafel1}.  The estimate of $\lambda_2$ (see figure \ref{F5}) is obtained 
by determining the maximum in the finite size approximation
\begin{equation}
\log C_w(\lambda) \approx \Sfrac{1}{n} \log c_n(\lfl \lambda n \rfl) .
\end{equation}
In the square lattice these estimates were extrapolated in the same way as 
in figure \ref{f9} gives, for example, $\lambda_2 = 0.723(5)$ when $w=1$, 
and $\lambda_2=0.624(2)$ if $w=2$.  In the cubic lattice more care was needed.
The estimates were distributed along a curve when plotted against $1/n$, and it
proved difficult to extrapolate it.  Experimentation shows that the data lines
up along a straight line segment when plotted against $(\log n) / \sqrt{n}$.
Extrapolating this gives the estimates in table \ref{tafel1}.  The results give
estimates close to zero, consistent with a conjecture that $\lambda_2=0$
in the cubic lattice.

\section{A self-avoiding walk underneath a piston}
\label{S3}

Now turn to the full model in figure \ref{f1}.  A walk from the origin
located on the anvil and centered underneath the piston is pushed
into a slit (in two dimensions) or a slab (in three dimensions) 
$S_w(\lambda)$ and may escape from underneath the piston at its edge.

The (finite size) free energy $\rho_n(\lambda) = (\log w_n(\lambda))/n$
is defined in equation \Ref{e1}, where (as in section \ref{s2}), $w_n(\lambda)$ 
is the number of self-avoiding walks from the origin of length $n$ avoiding 
the piston and the anvil.  The limiting free energy is defined in theorem
\ref{T1}. If $\lambda\geq 1$, then $\rho(\lambda) = \log \mu^{(d)}_w$ 
and the limit exists (equation \Ref{e3}) since the walk is entirely contained 
in the slit or slab above the anvil and underneath the piston.  More generally 
$\rho(\lambda) \geq \log \mu_w^{(d)}$ if $\lambda \leq 1$.

In general the walk is partitioned as illustrated in figure \ref{f3}, where a 
walk of length $n$ has its first section of length $\lfl \delta n \rfl$ confined
to the slit or slab $S_w(\lambda)$, until it first exits this region to become
a walk of length $n-\lfl \delta n\rfl$ into the bulk (this part of the walk 
may reenter and reexit the slit or slab $S_w(\lambda)$.

\subsection{The free energy $\rho(\lambda)$}

Lower and upper bounds on $w_n(\eps)$ will be obtained by either fixing
$x$ in figure \ref{f3} at height zero, or otherwise cutting the walk at
$x$ into a walk of span $\lfl \lambda n \rfl$ underneath the piston, and 
replacing its remaining part by an arbitrary self-avoiding walk.

\subsubsection{Lower bound:} Denote the number of walks in a $90^o$ wedge $W$, 
starting in the vertex of $W$ (square lattice), or in the spine (cubic)
of $W$, of length $n$, by $c_n^{\ra}$.  Then it is known that
$\lim_{n\to\infty} \Sfrac{1}{n} \log c_n^{\ra} = \log \mu_d$ \cite{HW85}.

If the vertex $x$ in figure \ref{f3} has height zero, then a lower bound 
is obtained by splitting the walk into a \textit{loop} of span $\lfl \lambda n \rfl$ 
from the origin to $x$ in $S_w(\lambda)$ underneath the piston.  The remaining
part of the walk, of length $n-\lfl\lambda n\rfl$, is confined to the $90^o$ wedge
$W$ formed by the edge of the piston and the anvil.  This walk starts in the
vertex $x$ located in the spine of the wedge.  Since this arrangement undercounts
$w_n(\lambda)$ (and $w_n(\lambda) \geq c_n^{(w)}$), it follows that
\begin{equation}
w_n(\lambda) \geq 
\ell_{\lfl \delta n\rfl}^{(w)}(\lfl\lambda n\rfl)\times c_{n-\lfl\delta n\rfl}^{\ra}.
\end{equation}
Take logarithms, divide by $n$ and let $n\to\infty$ to see that, by
theorems \ref{T2} and \ref{T4},
\begin{equation}
\liminf_{n\to\infty} \Sfrac{1}{n} \log w_n(\lambda)
\geq \max\{\log \mu^{(d)}_w\,,
\,\delta\log L_w(\lambda/\delta) + (1-\delta)\log \mu_{d} \} .
\end{equation}
The right hand side can be optimized by taking the supremum over $\delta$:
\begin{equation}
\liminf_{n\to\infty} \Sfrac{1}{n} \log w_n(\lambda)
\geq \max\{\log \mu^{(d)}_w\,,
\,\sup_{\lambda<\delta<1}(\delta \log L_w(\lambda/\delta) + (1-\delta)\log \mu_{d}) \} .
\label{e25}
\end{equation}

\subsubsection{Upper bound:} The situation is slightly different in two and in
three dimensions, so consider these in turn.

\textit{Two dimensions:} Cut the walk into two segments at the vertex $x$.
Replace the first segment of the walk with a walk of length 
$\lfl\epsilon n\rfl \in(\lfl \lambda n\rfl,n)$ and span $\lfl\lambda n\rfl$ 
in a slit of width $w$.  The remaining segment of the walk is replaced by an 
arbitrary self-avoiding walk of lenth $n-\lfl \epsilon n \rfl$ from $x$.  This gives an 
upper bound on the number of walks from the origin underneath the piston with 
the first $\lfl \epsilon n \rfl$ steps to $x$ inside the slit:
\begin{equation}
c_{\lfl \epsilon n \rfl}^{(w)}(\lfl \lambda n \rfl)\times c_{n-\lfl \epsilon n\rfl} .
\end{equation}
For each value of $n$, there exists an $\epsilon\in [\lambda, 1]$ maximizing 
this product.  Denote this value by $\epsilon^*$ (a function of $n$). Define
$k_n = \lfl \epsilon^* n\rfl$.  Then, for each $n$, define
\begin{equation}
v_n(\lambda) =  c_{k_n}^{(w)}(\lfl \lambda n \rfl)\times c_{n-k_n} .
\label{e27}
\end{equation}

Next, consider equation \Ref{e27},  take logarithms, divide by $n$, and take 
the limit superior on the right hand side as $n\to\infty$.  Then $n\to\infty$ on 
the right hand side along a subsequence $(n_i)$, and since $(k_n/n)$ is 
bounded, it has an accumulation point $\delta\in [\lambda,1]$ along 
a subsequence $(n_{ij})$ so that $(k_{n_{ij}}/n_{ij}) \to \delta$ 
as $j\to\infty$.  Put $m_j = n_{ij}$ and $r_j = k_{n_{ij}}$.  Then $(m_j)$ 
is a subsequence of $(n_i)$ and $(r_j)$ is a subsequence of
$(k_{n_i})$, such that $(r_j/m_j) \to\delta$ as $j\to\infty$.

That is, $(m_j) \subseteq (n_i)$ and $(r_j)\subseteq (k_{n_i})$.

Taking the limit superior of the right hand side of equation \Ref{e27} along
the subsequence $(m_j)$ then gives
\begin{equation}
\hspace{-2cm}
\Sfrac{1}{n} \log v_n(\lambda) 
\leq \limsup_{n\to\infty} \Sfrac{1}{n} 
   \log \LB c_{k_n}^{(w)}(\lfl \lambda n \rfl)\times c_{n-k_n} \RB
= \lim_{j\to\infty} \Sfrac{1}{m_j}
   \log \LB c_{r_j}^{(w)}(\lfl \lambda m_j \rfl)\times c_{m_j-r_j}\RB .
\end{equation}
Since $(r_j/m_j) \to \delta$ as $j\to\infty$, the result is that
\begin{equation}
\Sfrac{1}{n} \log v_n(\lambda) 
\leq  \kappa \log C_w(\lambda/\delta) + (1{-}\delta)\log \mu_2 .
\end{equation}
Taking the supremum over $\delta$ on the right hand side, and then 
the limit superior of the left hand side as $n\to\infty$ gives
\begin{equation}
\limsup_{n\to\infty} \Sfrac{1}{n} \log v_n(\lambda) 
\leq  \sup_{\lambda\leq\delta\leq 1} 
\LB \kappa \log C_w(\lambda/\delta) + (1{-}\delta)\log \mu_2  \RB.
\label{e30}
\end{equation}

It is also the case that 
\begin{equation}
w_n(\lambda) \leq \max\{ c_n^{(w)}, v_n(\lambda) \}
\label{e28}
\end{equation}
for each $n$.
Thus, by equation \Ref{e30}, 
\begin{equation}
\limsup_{n\to\infty} \Sfrac{1}{n} \log w_n(\lambda) 
\leq \max \{ \log \mu_w^{(2)} ,
\sup_{\lambda<\delta<1} 
(\delta \log C_w(\lambda/\delta) + (1-\delta)\log \mu_2) \} .
\label{e31}
\end{equation}

\textit{Three dimensions:}  This is analysed similarly to the two dimensional case,
but with the complication that the location of $x$ in figure \ref{f3} also 
depends on the shape of the piston.  In particular, projecting the piston
onto the anvil, the image of $x$ is located in the boundary of the projected
piston.  The shortest self-avoiding walk from the origin to such a vertex $x$
is $\alpha\lfl \lambda n \rfl+m$ where $\alpha\geq 1$ and $m\in [0,w-1]$.  
For example, if the piston is square, then $1\leq \alpha \leq 2$, and if it is 
circular, then $1\leq \alpha \leq \sqrt{2}$. Notice that $\lambda\,\alpha \leq 1$.

In general, for given $\lambda$ and $\delta$ in figure \ref{f3},
$\alpha$ is also a function of $n$, so write $\alpha=\alpha(n)$.  As $n\to\infty$,
$\alpha(n)$ varies, so define $\alpha^* = \liminf_{n\to\infty} \alpha(n) \geq 1$.   
Then $\lambda\,\alpha^*\leq 1$.

Proceeding as in the two dimensional case gives equation \Ref{e30} but with 
$\kappa \in [\eps\,\alpha^*,1]$. Since $\alpha^*\geq 1$, extend the range of
$\delta$ in the supremum to $(\lambda,1)$ to recover equation \Ref{e31}.

By theorem \ref{T6}, and equations \Ref{e25} and \Ref{e31}, existence of the
free energy is obtained for $\lambda\in[\lambda_1,1]$:

\begin{theorem}
If $\lambda \in [\lambda_1,1]$, then\\
$$ \rho(\lambda) 
= \lim_{n\to\infty} \Sfrac{1}{n} \log w_n(\lambda)
= \max\{\log \mu^{(d)}_w,
\,\sup_{\lambda<\delta<1}(\delta \log L_w(\lambda/\delta) + (1-\delta)\log \mu_{d}) \}.
\eqno \qed $$
\label{T7}
\end{theorem}

This also completes the proof of theorem \ref{T1}.  We conjecture that theorem 
\ref{T7} is valid for all $\lambda \in [\lambda_0,1]$.

By theorem \ref{T7} the following bounds can be obtained on $\rho(\lambda)$:
\begin{equation}
\max ( \log \mu_w^{(d)},(1-\lambda)\log \mu_d ) \leq \rho(\lambda)
\leq \log \mu_d + \lambda \log(\mu_w^{(d)}/\mu_d) . 
\end{equation}
These bounds follow since $0 \leq \log C_w(\lambda/\delta) \leq \log \mu_w^{(d)}$
for all $\lambda\in [0,1]$ and $\delta \in [\lambda,1]$.

\subsection{The critical point $\lambda_c$}

By theorem \ref{T7} 
\begin{equation}
\rho (\lambda) =  \max\{\log \mu^{(d)}_w,
\sup_{\lambda<\delta<1}(\delta \log L_w(\lambda/\delta) + (1-\delta)\log \mu_{d}) \} ,
\quad\hbox{for $\lambda\in[\lambda_1,1]$}.
\label{e35}
\end{equation}
Since $\rho(\lambda)$ is a constant for large $\lambda$, and a function of 
$\lambda$ for small $\lambda$, it is a non-analytic function of $\lambda$
and is singular at least at one point $\lambda_c$ defined by
\begin{equation}
\lambda_c = \inf\{\lambda \svv \rho(\lambda) = \log \mu_w^{(d)} \}.
\end{equation}
By noting that $0\leq \log C_w(\lambda/\delta)\leq \log \mu_w^{(d)}$, 
a bound on the location of $\lambda_c$ can be determined. Replacing 
$\log C_w(\lambda.\delta) = 0$ shows that if $\delta$ is small enough that
$(1-\delta)\log \mu_{d} > \log \mu_w^{(d)}$, then $\rho(\lambda)>\log \mu_w^{(d)}$.
This can only occur if $\delta$ (and thus $\lambda$) is small enough, giving
an lower bound on $\lambda_c$:
\begin{equation}
\lambda_c \geq \lambda_* = \frac{\log (\mu_d/\mu_w^{(d)})}{\log \mu_d} .
\label{44}
\end{equation}
If $\lambda > \lambda_c$, then the self-avoiding walk does not escape from the
region $S_w(\lambda)$ underneath the piston, and if $\lambda < \lambda_*$,
then the self-avoiding walk escapes from $S_w(\lambda)$.  Notice that it is 
not obvious that $\lambda_c\geq \lambda_1$.

If $d=2$ and $w=0$, then $\mu_0^{(2)}=1$ so that $\lambda_c=1$.  This
shows that for all $\lambda<1$ the walk escapes, as one expects in this case.
If $w=1$ instead, then $\mu_1^{(2)}=(1+\sqrt{5})/2$ (Sloane online sequences
of integers A038577; see also reference \cite{N12}), so that, 
using the estimate of $\mu_2$ in equation \Ref{2}, and the data in 
table \ref{tafel1}, $\lambda_c \geq \lambda_* = 0.50394\ldots$. This is the
lower bound in table \ref{tafel2} for $w=1$ in the square lattice.
The remaining bounds $\lambda_*$ are obtained by using the data in
table \ref{tafel1}.  In the cubic lattice, the estimate of the growth constant is 
given in equation \Ref{2}, and with the data in table \ref{tafel1} gives 
the estimate of $\lambda_*$ in table \ref{tafel2}.

To estimate the value of $\lambda_c$ numerically, consider equation \Ref{e35}
and define, for $\lambda\in [0,1]$ and $\lambda \leq \delta \leq 1$,
\begin{equation}
Q(\lambda,\delta) = \delta \log L_w(\lambda/\delta) + (1-\delta)\log \mu_{d}
- \log \mu^{(d)}_w .
\end{equation}
For large $\lambda$, the walk is confined under the piston, so that its
free energy is equal to $\log \mu^{(d)}_w$.  That is, for all $\delta \in[\lambda,1]$,
$Q(\lambda,\delta) \leq 0$.  On the other hand, if $\lambda$ is small, then 
the walk escapes.  In this case there must exist a $\delta \in[\lambda,1]$ such
that $Q(\lambda,\delta)>0$.  

Using the finite size approximation, one may consider
\begin{equation}
Q_n(\lambda,\delta) = \delta (\Sfrac{1}{n} \log c_n^{(w)}(\lfl \lambda n/\delta\rfl) 
+ (1-\delta)\log \mu_{d}) - \log \mu^{(d)}_w
\label{42}
\end{equation}
and find the largest value of $\lambda$ (say $\lambda_n$) such that there 
exists a $\delta\in[\lambda_n,1]$ such that $Q_n(\lambda_n,\delta)>0$. Then 
$\lambda_n$ is an approximation of $\lambda_c$.  Increasing $n$ gives a 
sequence $(\lambda_n)$ which can be extrapolated to $n=\infty$ to obtain a 
limiting estimate.  In the case that $w=0$ in the square lattice, this
trivially gives $\lambda_n = 1$, consistent with the critical point at $\lambda_c=1$.

\def\0{\phantom{$-$}}

\begin{table}[t!]
\caption{Numerical estimates of $\lambda_c$}
\begin{indented}
\lineup
\item[]
\begin{tabular}{llll}
\br                              
$w$ & dimension & $\lambda_*$ (lower bound)$\dagger$ & $\lambda_c$ (best estimate)  \cr 
\mr
$0$ & \multirow{6}{*}{square lattice} & $1$ & $1.0$  \cr
$1$ & & $0.5039\ldots$            & $0.758(3)$   \cr
$2$ & & $0.3305\ldots$            & $0.663(5)$        \cr
$3$ & & $0.2415\ldots$            & $0.612(2)$        \cr
$4$ & & $0.1876\ldots$            & $0.569(8)$        \cr
$5$ & & $0.1517\ldots$            & $0.540(8)$        \cr
\mr
$0$ & \multirow{6}{*}{cubic lattice} & $0.3717\ldots$ & $0.522(2)$  \cr
$1$ & & $0.1789\ldots$ & $0.360(2)$ \cr
$2$ & & $0.1111\ldots$ & $0.292(3)$ \cr
$3$ & & $0.0772\ldots$ & $0.251(3)$ \cr
$4$ & & $0.0574\ldots$ & $0.222(4)$ \cr
$5$ & & $0.0447\ldots$ & $0.201(4)$ \cr
\br
\end{tabular}
\begin{tablenotes}
\item $\dagger$ Equation \Ref{44}. 
\end{tablenotes}
\end{indented}
\label{tafel2}   
\end{table}

\begin{figure}[t]
\centering
\includegraphics[width=0.9\textwidth]{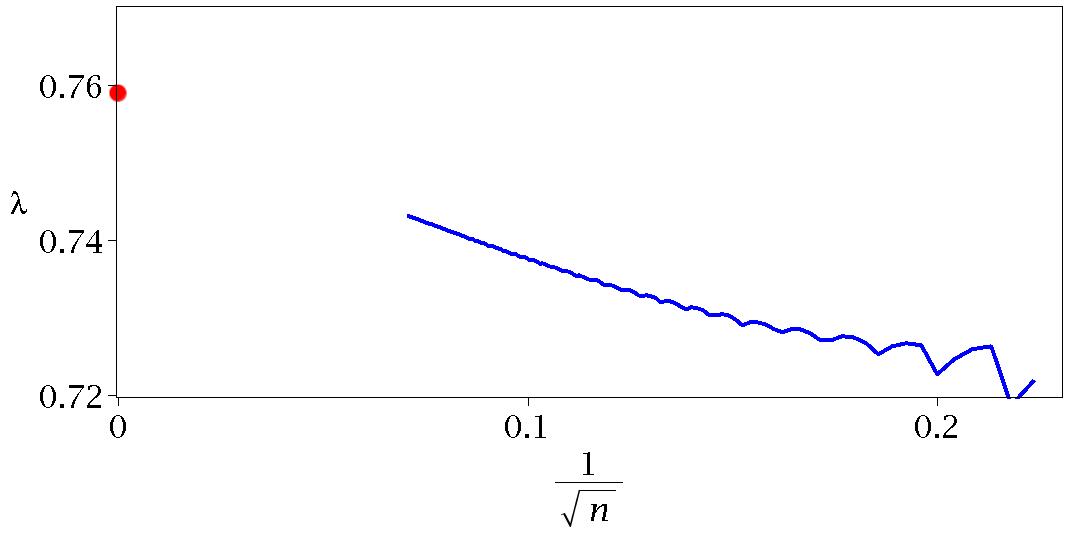}
\caption{Finite size approximations to $\lambda_c$ for $10\leq n\leq200$
in the square lattice and for $w=1$.  The initial partiy effects die down as
$n$ increases.  Extrapolating the curve to the $y$-axis using a quadratic 
curve gives the estimate $\lambda_c \approx 0.759$.}
\label{f10}
\end{figure}

In the case that $w>0$ in the square lattice, the data collected in section
\ref{S213} can be used to estimate $\lambda_n$, and then extrapolate it to
$\lambda_c$.  If $w=1$ in the square lattice then by equation \Ref{42}
estimates of $\lambda_n$ can be obtained and plotted as a function of $1/n$.
The resulting graph curves as $n$ increases, but plotting against $1/\sqrt{n}$ straightens
the curve somewhat (figure \ref{f10}).  Extrapolating this using a quadratic
curve gives $\lambda_n \to \lambda_c \approx 0.758$ as $n\to\infty$.  This
estimate is shown in table \ref{tafel2}.  The error bar is obtained by
using a linear extrapolation and comparing results.  The absolute difference
between the estimates is taken as a confidence interval.

\begin{figure}[t]
\centering
\includegraphics[width=0.9\textwidth]{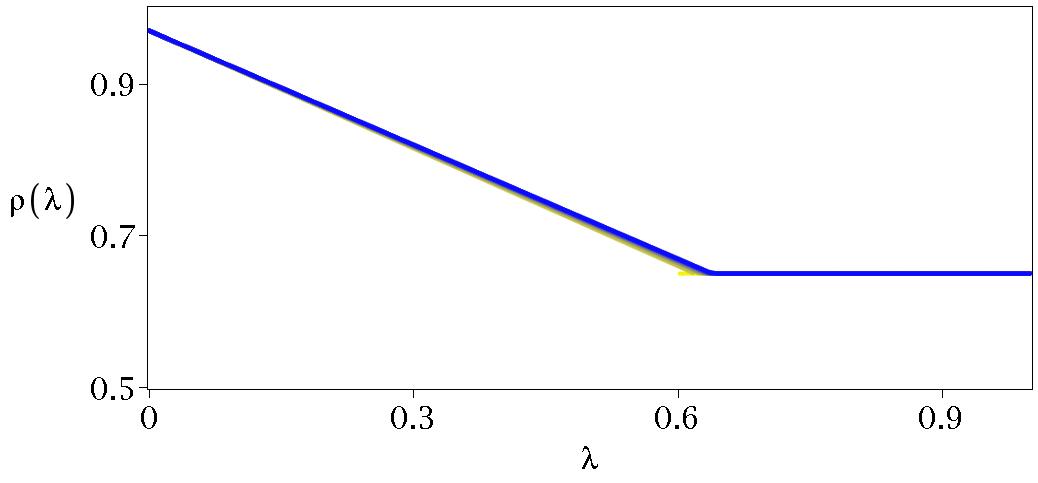}
\caption{Finite size approximations of the free energy $\rho(\lambda)$
for $w=2$ in the square lattice.  The critical point is well defined at
$\lambda_c \approx 0.663$ (see table \ref{tafel2}.  If $\lambda<\lambda_c$, then
the radius of the piston is small and the walk escapes so that $\rho(\lambda)$ is
not constant.  If $\lambda>\lambda_c$, then the radius of the piston is
large, and the walk is retracted underneath the piston so that it is not a function
of $\lambda$.  These plots are for $n=10,20,\ldots,200$ and they lie very close 
to each other.  The sharp corner at the critical point suggests a strong 
first order transition.}
\label{f11}
\end{figure}

Graphs of the finite size free energy can similarly be obtained from the
numerical data.  The finite size numerical approximation is obtained from
$Q_n(\lambda,\delta)$ (equation \Ref{42}) by
\begin{equation}
\rho(\lambda) \approx 
\max\LB \max_{\lambda \leq \delta\leq 1}\LB Q_n(\lambda,\delta) \RB 
, \log \mu_w^{(2)} \RB .
\end{equation}
In figure \ref{f11} our data is plotted for $w=2$ in the square lattice
for $n = 10,20,\ldots, 200$. The graphs are very close to each other
and show two regimes, namely a retracted phase whent he walk is confined
to the slit for large $\lambda$ and the free energy is independent of
$\lambda$, and an escaped phase where the walk escapes and the
free energy is a function of $\lambda$.  For $\lambda<\lambda_c$ the
free energy is dependent on $\lambda$, and for
$\lambda > \lambda_c$ by $\log \mu_w^{(2)}$.  The sharp transition
at the critical point is consistent with a strong first order transition
as the polymer escapes from underneath the piston.

\section{Conclusions}

Our numerical analysis show a very clear phase transition separating a retracted phase
and an escaped phase in the model.  The data indicate a strong first order 
transition at a critical point $\lambda_c$ which was estimated numerically and
listed in table \ref{tafel2}. 

The existence of a critical point was proven in section \ref{S3}, relying on
the results for ballistic walks in slits and slabs in section \ref{S2}.  The analysis
of these ballistic walks proved existence of a limiting free energy in the model
in figure \ref{f1} for a range of values of $\lambda$ as stated in theorem
\ref{T1}.  However, there remains an interval $[\lambda_0,\lambda_1)$
(corollary \ref{cor4} and corollary \ref{cor5}) where
existence of the free energy was not proven.  We conjectured in both cases
that $\lambda_1^\prime=\lambda_1=\lambda_0$, so that the free energy
exists for all $\lambda\in[\lambda_0,1]$, but this remains to be proven.

Our main results are summarized in theorem \ref{T1} and in theorem \ref{T7}
(where an explicit expression is given for the free energy $\rho(\lambda)$ in
terms of the growth constant $L_w(\lambda)$ of ballistic loops in a slit or
in a slab in the square and cubic lattices respectively).  However, the lower bound
in equation \Ref{44} proved to be quite weak, as seen in table \ref{tafel2}.

\section*{Acknowledgement}
EJJvR acknowledges financial support from NSERC (Canada) 
in the form of a Discovery Grant RGPIN-2019-06303 and is grateful to SG
Whittington for discussions about this model.   Data generated for this study
are available on reasonable request.

\section*{References}
\bibliographystyle{plain}
\bibliography{piston.bib}

\end{document}